\documentclass{article}

\usepackage{hyperref}       
\usepackage{url}            
\usepackage{booktabs}       
\usepackage{amsfonts}       
\usepackage{nicefrac}       
\usepackage{microtype}      
\usepackage{lipsum}		
\usepackage{graphicx}
\usepackage{doi}
\usepackage{enumitem}
\usepackage{amsthm}

\usepackage{booktabs}
\usepackage{amssymb}
\usepackage{amsmath}
\usepackage{stmaryrd}
\usepackage{mathrsfs}
\usepackage{color}
\usepackage{graphicx}
\usepackage{latexsym}
\usepackage{tabls}
\usepackage{graphicx}
\usepackage{epsfig}
\usepackage{subfigure}
\usepackage{rotating}
\usepackage{latexsym}
\usepackage[mathcal]{eucal}
\usepackage{amssymb}
\usepackage{colortbl}
\usepackage{times}
\usepackage{longtable}
\usepackage{algorithmic}
\usepackage{algorithm}
\usepackage{psfrag}
\usepackage{tabularx}
\usepackage{epstopdf}
\usepackage{gensymb}
\usepackage{bbm}
\usepackage{setspace}
\usepackage{fullpage}
\usepackage{url}
\usepackage{calc}
\usepackage{siunitx}

\usepackage{lineno,hyperref}
\usepackage{stmaryrd}
\usepackage{subfigure}
\modulolinenumbers[50]
\usepackage{epstopdf}
\usepackage{array}
\usepackage{textcomp}

\newcommand{\opA}{\mathop{\vphantom{\sum}\mathchoice
  {\vcenter{\hbox{\huge A}}}
  {\vcenter{\hbox{\Large A}}}{\mathrm{A}}{\mathrm{A}}}\displaylimits}
  
\usepackage{booktabs}
\usepackage{mathrsfs}
\usepackage{multirow}
\usepackage{lscape}
\usepackage{amsmath}

\usepackage{mathtools}
\usepackage{amssymb}
\usepackage{color,soul}
\usepackage{bm}

\usepackage{arxiv}

\usepackage[utf8]{inputenc} 
\usepackage[T1]{fontenc}    
\usepackage{hyperref}       
\usepackage{url}            
\usepackage{booktabs}       
\usepackage{amsfonts}       
\usepackage{nicefrac}       
\usepackage{microtype}      
\usepackage{lipsum}
\usepackage{graphicx}
\usepackage{amsmath}
\usepackage{bm}
\graphicspath{ {./images/} }

\title{Configurational forces explain echelon cracks in soft materials}

\author{
\href{https://orcid.org/0000-0002-7898-2986}{\includegraphics[scale=0.06]{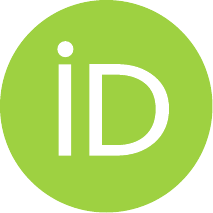}\hspace{1mm}Angel Santarossa}
\thanks{Corresponding author: angel.santarossa@fau.de} \\
  Institute of Applied Mechanics\\
  Friedrich-Alexander-Universität Erlangen–Nürnberg\\
  Egerlandstr. 5, 91058, Erlangen, Germany \\
  \texttt{angel.santarossa@fau.de} \\
   \And
 \href{https://orcid.org/0000-0002-3703-6117}{\includegraphics[scale=0.06]{orcid.pdf} Nydia Roxana Varela-Rosales} \\
  Department of Complex and Intelligent Systems\\
  Future University Hakodate\\
  041-8655, Hokkaido, Japan \\
  \texttt{nydia@fun.ac.jp} \\
  \And
 \href{https://orcid.org/0000-0003-1490-947X}{\includegraphics[scale=0.06]{orcid.pdf} Paul Steinmann} \\
  Institute of Applied Mechanics\\
  Friedrich-Alexander-Universität Erlangen–Nürnberg\\
  Egerlandstr. 5, 91058, Erlangen, Germany \\
  Glasgow Computational Engineering Centre\\
  School of Engineering, University of Glasgow\\
  \texttt{paul.steinmann@fau.de} \\
\And
\href{https://orcid.org/0000-0002-3476-2180}{\includegraphics[scale=0.06]{orcid.pdf}\hspace{1mm}Miguel Angel Moreno-Mateos}\thanks{Corresponding author: miguel.moreno@fau.de} \\
  Institute of Applied Mechanics\\
  Friedrich-Alexander-Universität Erlangen–Nürnberg\\
  Egerlandstr. 5, 91058, Erlangen, Germany \\
  \texttt{miguel.moreno@fau.de} \\
}

\begin{document}
\maketitle
\begin{abstract}
Soft fracture in highly deformable solids involves both geometric and constitutive nonlinearities, necessitating advanced theoretical and computational frameworks for its accurate understanding. Tensile fractures subjected to mixed-mode loading deviate from their original planar shape, resulting in echelon crack patterns. When out-of-plane shear is superimposed, a crack front segments into an array of tilted facets. The physical interpretation of echelon cracks is only marginally understood, and it is customarily based on rather limited approaches based on Linear Elastic Fracture Mechanics. Here we investigate mixed-mode I + III fracture within the framework of configurational mechanics. Using the Configurational Force Method, implemented as a post-processing algorithm in a finite-element-based simulation, we compute the configurational forces acting at the crack tip of model fracture geometries prior to propagation. Configurational forces characterize both the magnitude and direction of propagation for maximal energy release rate. Our results reveal the complex interactions between tilted facets and their critical role in shaping the fracture morphology. We also examine the effects of facet coalescence—driven by the growth of the parent crack—where neighboring facets merge into a unified crack front. These findings provide new insights into fracture processes in soft, quasi-brittle materials under mixed-mode loading. 
\end{abstract}


\section{Introduction}

From engineering failures to geological formations, crack propagation governs material integrity across scales. Despite decades of research, predicting and controlling fracture remains a major challenge in mechanics. Quasi-static planar crack growth under tensile stress normal to the fracture plane is well understood; however, under varying conditions, cracks can develop geometrically complex patterns. For instance, when a crack exceeds a critical velocity—such as a finite fraction of the Rayleigh wave speed in brittle materials—it can transition from smooth propagation to dynamic instability, often causing surface roughening \cite{Buehler2006, RaviChandar1998, Fineberg1991, Kolvin2015, GoldmanBoue2015b}. In slow fractures, complex fracture geometries, such as step-like patterns, can emerge due to local distortions of the stress field induced by material heterogeneity \cite{Wu2025, Steinhardt2022, Steinhardt2023, Lechenault2015, Ramanathan1997}.

Loading variations can induce deviations in the crack path, even in homogeneous materials. Under pure uniaxial tension, cracks typically propagate perpendicular to the applied load (Mode I). However, when shear stresses are introduced, crack trajectories become more complex. In-plane shear (Mode II) causes the entire crack front to kink and propagate in a tilted direction. In contrast, out-of-plane shear (Mode III) induces the crack front to transition into an array of tilted facets—known as echelon patterns. Interestingly, this complex 3D fracture morphology emerges across diverse materials, from soft (e.g., gels~\cite{Pham2016,Ronsin2014} and cheese~\cite{Goldstein2012, Goldstein2014}) to hard solids (e.g., rocks~\cite{Pollard1982}, glass~\cite{SOMMER1969539} and polymers~\cite{Lazarus2008, Lin2010}). 

\begin{figure}[htb]
\centering
	\includegraphics[width=0.9\columnwidth]{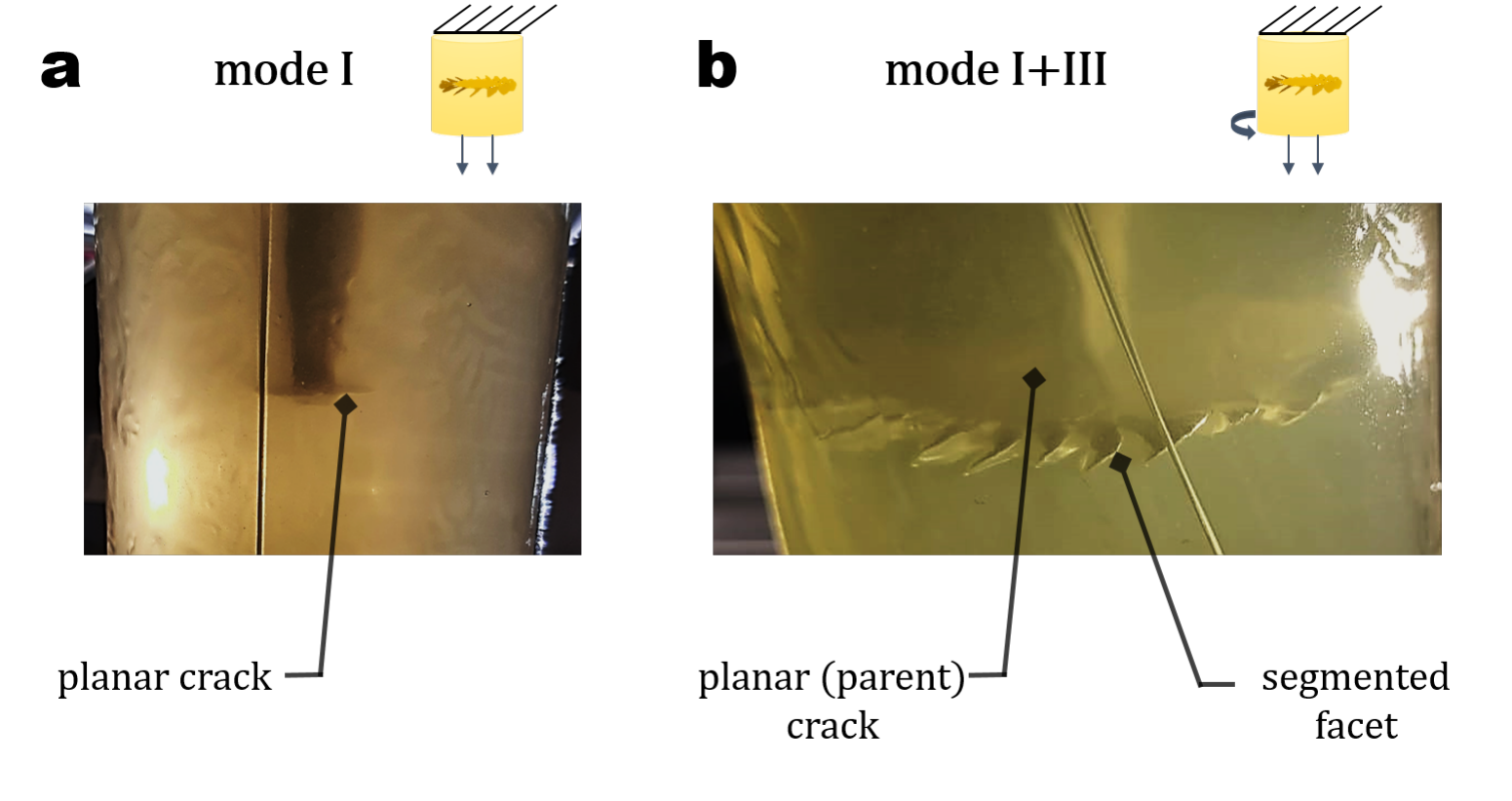}
    \caption{\label{fig:example_mixed_mode} \textbf{Crack front segmentation under mixed-mode I+III loading in a hydrogel.} (a) A planar crack is initiated under uniaxial tensile loading (Mode I) and propagates as tension increases. (b) Upon superimposing torsion (Mode III), the initially planar crack front segments into tilted facets. The visible line along the hydrogel surface results from the molding process and is unrelated to the fracture. Schematic insets illustrate the loading conditions: the hydrogel is fixed at the top, while tension and counterclockwise torsion are applied at the bottom.}
\end{figure}

Traditionally, crack propagation has been studied using Linear Elastic Fracture Mechanics (LEFM). The theory assumes ideal brittle, linear-elastic materials undergoing small deformations. Nonlinear effects are confined to a singular region near the crack front, and the stress near the crack tip exhibits a characteristic square root dependence on the distance from the crack tip. This stress field behavior is further characterized by the stress intensity factors $K_\text{I}$, $K_\text{II}$, and $K_\text{III}$ for modes I, II, and III, respectively. These factors are essential for quantifying the stresses that develop under various loading conditions. Echelon cracks have been extensively studied within the framework of LEFM \cite{Pham2014, Pham2016, Wu2006, ortellado2025principle, LEBLOND2011, CAMBONIE2014, Vasudevan2018, Lebihain2022, VASUDEVAN2020, LEBLOND2019187, BAHMANI2021}.

To predict crack nucleation and the direction of propagation, LEFM is complemented by empirical failure criteria. In particular, for mixed-mode I+III loading, a broadly accepted criterion is the Principle of Local Symmetry (PLS). It states that a crack will propagate in a direction where the shear stresses at the crack front vanish ($K_\text{II}=0$ and $K_\text{III}=0$) \cite{Lin2010}. Three-dimensional calculations suggest that quasi-static cracks are governed by the PLS, beyond a transitional region determined by microscopic scales near the crack front \cite{Hodgdon1993}. Notably, recent experiments on mixed-mode I+III fracture in nearly linear-elastic brittle gel demonstrated the validity of the PLS in mixed-mode fracture \cite{ortellado2025principle}. Deviations from the PLS in these experiments were attributed to elastic interactions between neighboring facets. The geometric configuration of the facets can either shield or amplify stresses near the crack front, leading to facet arrest and coarsening, which determine the final fracture geometry \cite{ortellado2025principle, Pham2016, Molnar2024}. Despite the critical role of facet interactions in crack segmentation and the resulting fracture morphology, the underlying physical mechanisms remain elusive and have not been thoroughly studied.   

In addition to the significant effects of facet interactions on crack propagation, material properties can also lead to substantial deviations from LEFM predictions. This is particularly evident in soft material fracture, where LEFM solutions lose validity due to their restrictive assumption of infinitesimal deformation \cite{Bouchbinder2009, Bouchbinder2009b, Bouchbinder2008, GoldmanBoue2015, Livne2010, Livne2008}. In this sense, Ronsin et al. \cite{Ronsin2014} studied the formation of echelon cracks in strain-hardening gels subjected to combined tensile and antiplane shear loading. They found that crack fragmentation is governed by the material's nonlinear elastic behavior, which LEFM cannot capture. Notably, deviations from LEFM arise even in the fracture of brittle soft materials under simple loading conditions. Recent experiments have shown that, in brittle, soft solids subjected to uniaxial tension, the crack front is not planar—a primary assumption in LEFM. Instead, it evolves into a complex geometry that breaks planar symmetry, thereby enhancing material toughness \cite{Wei2024}. These findings challenge the core assumptions of LEFM and highlight the need for more general theoretical frameworks, beyond linear elasticity, to accurately describe fracture behavior, particularly in soft materials.

Various approaches have been utilized to overcome some of the limitations of LEFM. In the context of crack segmentation due to mixed-mode loading, these include phase-field models~\cite{Pons2010,Chen2015,Pham2016, Henry_2016, Molnar2024}, cohesive zone models~\cite{Leblond2015,Lazarus2020, Hattali2021}, and eigenerosion~\cite{Pandolfi2012}. These frameworks typically rely on additional assumptions (e.g., specific parameters or criteria) or require regularization schemes to model the fracture process accurately. Significant model adjustments are often also needed when applied to materials with different mechanical responses.

Building on the need for a more general fracture framework, Configurational Mechanics (CM) offers a powerful alternative. In CM, the celebrated $J$-integral is calculated from the Eshelby stress tensor--configurational stress measure--\cite{Cherepanov1967, Rice1968} for any linear or nonlinear material behavior. The seminal computations of $J$ were \textit{via} a contour integral around the crack tip. Alternatively, the Configurational Force Method (CFM) proposes to directly discretize the Eshelby stress into nodal forces on the FE mesh. One key advantage of the CFM is its ease of implementation as a postprocessing algorithm, which involves i) calculating the Eshelby stress and ii) calculating the nodal forces. As mentioned, an overarching advantage of CM is its applicability to a broad range of material models without imposing specific assumptions. This includes materials exhibiting geometric and constitutive nonlinearities \cite{Schmitz2023}. Unlike LEFM, CM is not limited by assumptions of infinitesimal strains or small-scale yielding. This allows CM to capture the full complexity of crack tip behavior, including crack front instabilities and intricate fracture morphologies. It makes CM particularly well-suited for studying crack segmentation, facet formation, and other phenomena that involve large deformations or nonlinear material responses.

Of particular relevance to the present work, we earlier showed that configurational forces provide an accurate estimate of the $J$-integral in soft solids \cite{Moreno-Mateos2024b}. Configurational forces have also been calculated in strain gradient elasticity theories \cite{Serrao2025b}, biomechanics \cite{Goda2016}, FE remeshing \cite{Braun1997}, and crystal plasticity in metals, among other relevant fields. Overall, the CFM offers a more general approach for capturing the complex interactions at the crack tip, particularly under mixed-mode loading and finite strains in soft materials, which is critical to studying unstable crack growth such as fracture fragmentation.

Despite recent advances using phase-field and coupled-criterion models to investigate the formation and resulting morphology of segmented facets in echelon cracks---see the work by Molnár et al. \cite{Molnar2024}---, the role of interactions between facets remains only marginally understood. As Molnár et al. \cite{Molnar2024} and Ortellado et al. \cite{ortellado2025principle} underscore, configurational mechanics hold significant promise for shedding new light on these intricate interactions. In this work, we leverage configurational mechanics to study crack front segmentation in mixed-mode I+III fractures in a soft hydrogel. Our approach extends beyond the linear elastic assumptions of LEFM and enables us to capture the complex interactions at the crack tip. First, we develop an idealized geometrical model of cracks subjected to tensile and out-of-plane shear stress (i.e., echelon-like cracks), which replicates key geometrical features observed in earlier experiments—such as average facet number, tilting angle, facet spacing, and width-to-length ratio. Next, using the Configurational Force Method as a post-processing algorithm in a Finite Element (FE) simulation, alongside our simplified crack model, we compute configurational forces at the crack tip to investigate the underlying mechanism behind the fragmentation process. In particular, we study the intricate interactions among segmented facets and reveal their critical role in determining the overall fracture morphology. In addition to the interaction of individual facets, we also explore the phenomenon of facet coalescence, in which neighboring facets merge—driven by the growth of the parent crack—into a unified crack front. Although experimentally observed, this effect remains poorly understood. By applying a molecular dynamics-based mesh relaxation technique to our simplified geometry, we achieve a smooth representation of the junction between adjacent facets. This enables us to reveal how coalescence leads to asymmetric crack growth: it shields the propagation tendency in nearby facet regions while amplifying the driving forces in more distant parts of the coalescing front.

\section{Experimental observations and idealization of echelon cracks}
The calculation of crack tip configurational forces in this work is performed on FE meshes that idealize echelon-like crack patterns. We leverage experimental observations of echelon crack patterns to inform the development of an idealized crack geometry, which is implemented as an FE mesh. This mesh provides a simplified yet representative model to investigate configurational forces at the crack front.

By idealizing the geometry based on key experimental features, we reduce modeling complexity while preserving essential morphological parameters—such as the number of facets, tilt angle, and spacing. This approach enables a systematic investigation of facet interactions and their influence on the resulting fracture morphology.

\subsection{Review of empirical observations on echelon cracks}

Experiments show that even small contributions of anti-plane shear in an initial planar fracture can lead to the segmentation of the crack front \cite{SOMMER1969539, Pollard1982, Lazarus2008, Pham2014, Pham2016, Lin2010, Ronsin2014}. The fragmentation process occurs through the nucleation of discrete segments from the parent crack front. Generally, these daughter cracks do not form as smooth extensions of the parent fracture but rather emerge as distinct crack planes \cite{Pham2016,Pham2014,Lin2010, Goldstein2012, Goldstein2014}, except for significantly small mode III loading compared to mode I \cite{SOMMER1969539}. These facets, also usually termed type A cracks, appear as regularly spaced, unconnected segments \cite{Pham2016, Goldstein2012, Goldstein2014}, with their spacing observed to scale with the size of the parent crack \cite{Pham2014}.

Observations further indicate that the bridging region, i.e., the uncracked material between adjacent facets, remains intact in certain cases \cite{Chen2015, Pham2016, Lin2010, Goldstein2012, Goldstein2014, Lazarus2008}. In contrast, other experiments show that it fractures at later stages, allowing the parent crack front to advance toward the daughter cracks \cite{Pham2016, Lin2010, Pollard1982}. Theoretical calculations indicate that the fracture of the bridging region (type B cracks) is energetically less favorable than the propagation of type A cracks \cite{LAZARUS2001}.

The facets generally exhibit a characteristic finger-like shape, with curved tips and flat sides \cite{Chen2015, Lazarus2020, Hattali2021, Santarossa2023, ortellado2025principle}. Their width typically increases as they propagate. Furthermore, they are observed to align perpendicular to the local maximum tensile stress direction and maintain this orientation as they grow (unless additional mode II loading is imposed) \cite{Chen2015,SOMMER1969539, Pham2016}.

\begin{figure}[htbp]
  \centering
  \includegraphics[width=1.0\linewidth]{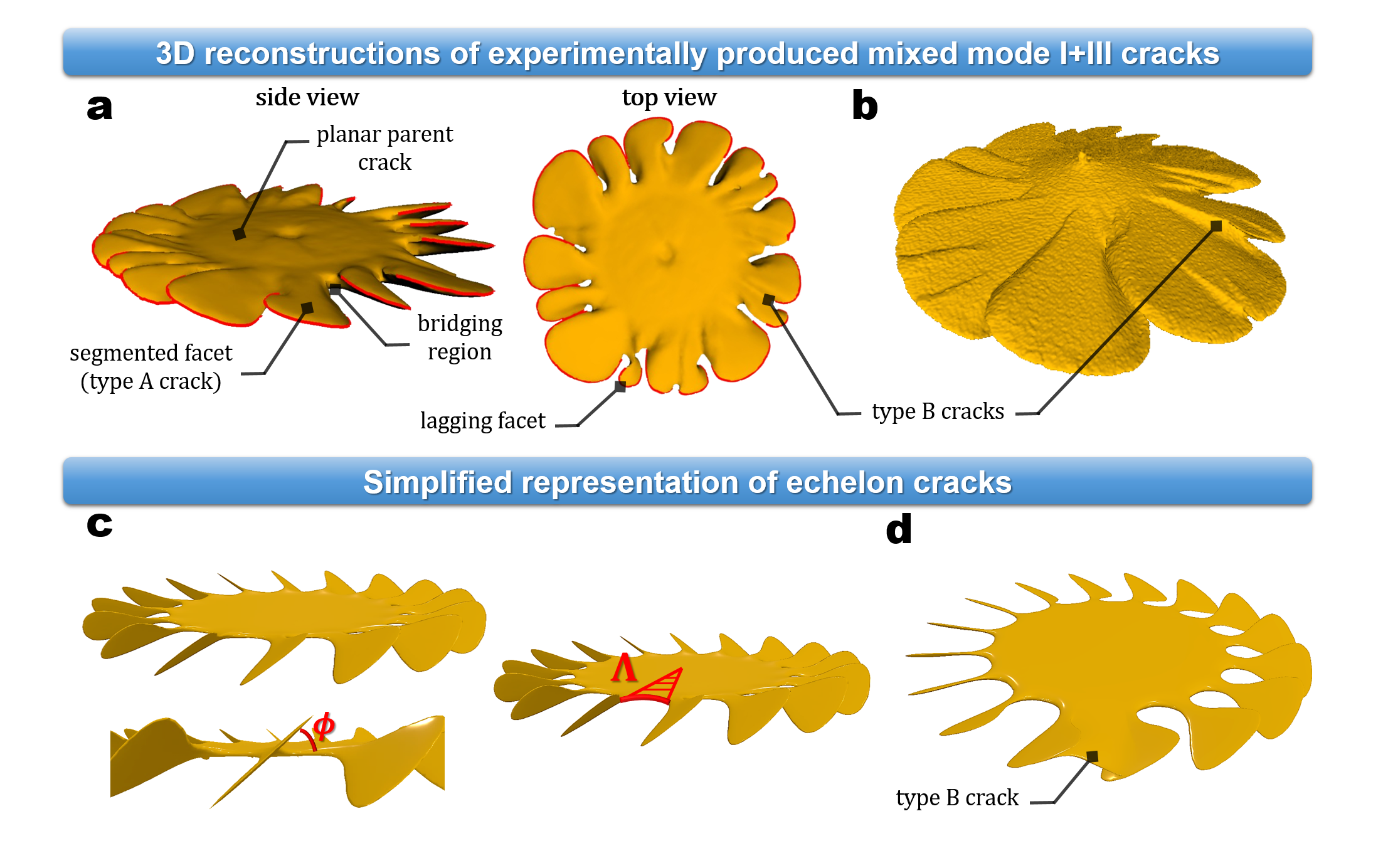}
  \caption{\textbf{Mixed-mode I+III crack morphologies and simplified echelon model representation.} Top panel: (a) and (b) show reconstructed echelon crack morphologies from X-ray tomography of cylindrical hydrogel samples fractured under mixed-mode I+III loading. The crack fronts display finger-like facets, curved tips, and type B bridging cracks. Note that samples were imaged under load, resulting in a visually thicker crack surface.  Images (a) and (b) are adapted from \cite{ortellado2025principle} and \cite{Santarossa2023}, respectively. 
  Bottom panel: proposed simplified representation of echelon patterns. An example of the idealized echelon crack model is shown in (c). The segmented facets are tilted an angle $\phi$ from the planar parent crack and symmetrically distributed with an angular spacing $\Lambda$. (d) illustrates a simplified model of a nucleated type B, where two neighboring facets have coalesced.}
  \label{fig:echelon_example}
\end{figure}

These characteristic features of echelon cracks are exemplified in Figures~\ref{fig:echelon_example}a and b, which show mixed-mode I+III cracks produced in cylindrical hydrogel samples reconstructed using X-ray imaging, as reported in \cite{ortellado2025principle, Santarossa2023}.

Uneven facet growth and the sudden arrest of propagating facets have also been observed in experiments \cite{Pham2016}. Furthermore, facets are observed to coarsen through the elimination of neighboring facets, resulting in increased facet width and spacing \cite{Chen2015}. The remaining facets maintain their original angle while expanding over others. Numerical studies suggest that these effects (facet arrest and coarsening) arise due to elastic interactions between adjacent facets \cite{Pham2016, Chen2015, ortellado2025principle}.

These experimental observations highlight the complexity of crack front segmentation, where multiple effects — such as facet coarsening, uneven growth, and bridging region fracture — often occur simultaneously. As a result, isolating individual mechanisms in a controlled and systematic manner in experiments is challenging. Simplified geometrical models provide a valuable tool to investigate these effects independently.

\subsection{Simplified geometrical representation of echelon cracks}

Building on these experimental observations, we propose a simplified geometrical model for echelon cracks.

Previous research has used circular cracks embedded in a rectangular planar crack as a simplified representation of mixed-mode I+III fractures \cite{ortellado2025principle, Pham2014, Pham2016}. Nonetheless, this geometrical representation does not effectively capture the finger-like shape with widening tips exhibited by segmented facets in experiments. An accurate model is essential to reliably investigate the complex interactions between tilted facets.

Inspired by experimental geometries reported by Santarossa et al. \cite{Santarossa2023} and Ortellado et al. \cite{ortellado2025principle}, in the present work, echelon cracks are represented by a central planar region (parent crack), modeled as a thin cylindrical plate with tilted segments (daughter cracks) extending from its edge, as illustrated in Fig~\ref{fig:echelon_example}c. The daughter cracks, or facets, are symmetrically arranged along the parent crack and evenly distributed at equal angular intervals, $\Lambda$ (Fig~\ref{fig:echelon_example}c). The facets are smoothly integrated into the parent crack, forming a continuous structure within the mesh. Each facet is inclined at a tilting angle $\phi$ (Fig~\ref{fig:echelon_example}c) relative to the plane of the parent crack. The profile of each facet is elongated, with a widening tip, resembling the characteristic finger-like shape observed in experiments \cite{ortellado2025principle, Chen2015, Lazarus2020, Hattali2021}. A detailed description of the mesh generation procedure is given in ~\ref{Model-crack-mesh-description}.

Fig.\ref{fig:echelon_example}d shows an extension of this model to represent facet coalescence, specifically through the development of a type B crack where initially segmented facets merge into a single front. To ensure a smooth geometric transition in the merging regions, we applied a molecular dynamics-based mesh relaxation technique, as described in\ref{Molecular-dynamics}. In this approach, mesh nodes are treated as particles interacting via a simple potential and undergo iterative energy minimization to reduce the system's total energy. This procedure reduces local distortion and produces a smooth surface at the union between facets.

In experiments, phenomena such as uneven facet sizes or the nucleation of type B cracks rarely occur in isolation, making it difficult to study them in a controlled manner. However, our approach enables a systematic investigation of crack segmentation.

\section{Continuum mechanics and configurational force method}

\subsection{Kinematics \& Field equations}
The deformation of the medium is formulated in a finite strain framework. The deformation field $\boldsymbol \varphi\left(\mathbf{X}\right)$ maps the positions in the material configuration $\mathbf{X} \in \Omega_0$, with boundary $\partial \Omega_0$, to the positions in the spatial configuration $\mathbf{x} \in \Omega$ according to $\mathbf{x}=\boldsymbol \varphi\left(\mathbf{X}\right)=\mathbf{u}\left(\mathbf{X}\right)+\mathbf{X}$, with $\mathbf{u}\left(\mathbf{X}\right)$ the displacement field. The deformation gradient is defined as $\mathbf{F}=\nabla_0\mathbf{u} + \mathbf{I}$, with $\mathbf{I}$ the second-order identity tensor and $\nabla_0$ the gradient operator in the material configuration. Following the multiplicative isochoric-volumetric decomposition $\mathbf{F}=\mathbf{F}_\text{vol}\cdot\overline{\mathbf{F}}$, the volumetric part is defined as $\mathbf{F}_\text{vol}=\left[ \det \mathbf{F}\right]^{1/3}\mathbf{I}$ and the isochoric part as $\overline{\mathbf{F}}=\left[\det\mathbf{F}\right]^{-1/3}\mathbf{F}$, where $\det \mathbf{F}$ denotes the determinant of $\mathbf{F}$.

The energy density per undeformed volume of the system consists of isochoric and volumetric contributions, according to a decoupled representation
\begin{equation}\label{eq:total_energy_density}
  \Psi(\mathbf{F}) = \Psi_\text{iso}(\overline{\mathbf{F}}) + \Psi_\text{vol}(\det \mathbf{F}).
\end{equation}

The potential energy functional $\Pi$ is defined as
\begin{equation}\label{eq:Potential_energy}
\Pi\left(\mathbf{u}\right) = \Pi_\text{int}\left(\mathbf{u}\right)+\Pi_\text{ext}\left(\mathbf{u}\right)=
\int_{\Omega_0} \Psi\left(\mathbf{F}\left(\mathbf{u}\right)\right)  \, \text{d}V 
- \int_{\Omega_0}  \mathbf{b}_0	\cdot \mathbf{u}   \, \text{d}V 
- \int_{\partial\Omega_0} \mathbf{t}_0\cdot\mathbf{u} \, \text{d}A,
\end{equation}
\noindent with $\mathbf{b}_0$ and $\mathbf{t}_0$ the body and surface forces, respectively.

The governing field equation can be obtained through the principle of stationary potential energy, which requires that the first variation of the total potential energy functional vanishes for any admissible virtual displacement $\delta\mathbf{u}$,
\begin{equation}\label{eq:Energy_functional}
\delta_{\mathbf{u}} \mathrm{\Pi} = \left.\frac{\mathrm{d}}{\mathrm{d} \lambda} \mathrm{\Pi}(\mathbf{u}+\lambda  \delta \mathbf{u})\right\vert_{\lambda=0} =
\int_{\Omega_0} \frac{\partial \Psi\left(\mathbf{F}\left(\mathbf{u}\right)\right)}{\partial \mathbf{F}} : \nabla_0 \delta\mathbf{u} \, \text{d}V 
- \int_{\Omega_0}  \mathbf{b}_0	\cdot \delta\mathbf{u}   \, \text{d}V 
- \int_{\partial\Omega_0} \mathbf{t}_0 \cdot \delta\mathbf{u} \, \text{d}A
\stackrel{!}{=} 0.
\end{equation}

From the previous requirement, in the absence of body and surface forces, and identifying the Piola stress tensor as $\mathbf{P}=\partial_\mathbf{F}\Psi\left(\mathbf{F}\left(\mathbf{u}\right)\right)$, the field equation reads
\begin{equation}\label{eq:strong_form}
\nabla_0 \cdot \mathbf{P} + \mathbf{b}_0 = \mathbf{0},
\end{equation}
\noindent with $ \nabla_0 \cdot$ the divergence operator in the material configuration. We note that in our computations body forces are omitted, thus the field equation simplifies to the zero divergence of the Piola stress tensor.

\subsection{Constitutive model}

The isochoric contribution in Equation~\ref{eq:total_energy_density} is here defined for a neo-Hookean material as a function of the isochoric deformation gradient,
\begin{equation}\label{eq:Psi_isochoric}
\Psi_\text{iso}\left(\overline{\mathbf{F}}\right)=\frac{G}{2} \left[ \mathbf{I} : \left[\overline{\mathbf{F}}^\text{T}\cdot\overline{\mathbf{F}}\right] - 3\right],
\end{equation}
\noindent where $G=9$~kPa is the shear modulus calibrated from tensile tests \ref{Model-calibration}.

The volumetric contribution is a function of the Jacobian of the deformation gradient,
\begin{equation}
\Psi_{\text{vol}} \left( \det \mathbf{F} \right) =
\frac{\kappa}{2} \left[\det\mathbf{F}-1\right]^2,
\end{equation}
\noindent with $\kappa$ the bulk modulus is obtained via $\kappa=\frac{2G\left[1+\nu\right]}{3\left[1-2\nu\right]}$. 
Hydrogels are typically considered incompressible, with a Poisson’s ratio of 0.5. However, van Otterloo and Cruden showed that the Poisson’s ratio decreases at high gelatin concentrations \cite{VANOTTERLOO201686}. For a hydrogel with a gel concentration of 10 wt.\%, we adopt a Poisson’s ratio of $\nu = 0.451$ based on their findings.

The Piola stress tensor can be derived from the energy density as the addition of isochoric and volumetric contributions, $\mathbf{P}=\mathbf{P}_\text{iso}+\mathbf{P}_\text{vol}$, where the isochoric contribution is obtained as

\begin{equation}\label{eq:Piola_stress}
\mathbf{P}_\text{iso} = \frac{\partial \Psi_\text{iso} \left(\overline{\mathbf{F}}\right)}{\partial \mathbf{F}} =
\left[\det\mathbf{F}\right]^{-1/3}\mathbb{K}:\frac{\partial \Psi_\text{iso}\left(\overline{\mathbf{F}}\right)}{\partial \overline{\mathbf{F}}}
= G \, \left[\det\mathbf{F}\right]^{-1/3}\mathbb{K} :  \overline{\textbf{F}}  ,
\end{equation}
\noindent with the fourth-order projection tensor $\mathbb{K}=\mathbb{I}-\frac{1}{3}\mathbf{F}^{-\text{T}}\otimes\mathbf{F}$, and the volumetric contribution,
\begin{align}\label{eq:Pvol_v}
\mathbf{P}_\text{vol}
=\frac{\partial \Psi_\text{vol}\left(\mathbf{F}\right)}{\partial \mathbf{F}}=\det\mathbf{F}\frac{\partial \Psi_\text{vol}\left(\mathbf{F}\right)}{\partial \det\mathbf{F}}\mathbf{F}^{-\mathrm{T}}
=\kappa\left[\left[\det\mathbf{F}\right]^2-\det\mathbf{F}\right]\mathbf{F}^{-\text{T}}.\nonumber
\end{align}

\subsection{Configurational Force Method}

Just as spatial forces are associated with the deformation of the continuum, configurational forces drive changes in the material configuration, leading to an associated release of energy. The definition of a stress-energy tensor in the material motion description is in the heart of the theory. Akin to the energy density $\Psi\left(\mathbf{F};\mathbf{X}\right)$ defined in $\Omega_0$, an energy density $\psi=\psi\left(\mathbf{f}, \boldsymbol \phi \right)$ can be re-defined per unit volume in $\Omega$, so that $\Psi = \left[\det \mathbf{F}\right] \, \psi$. This new energy density serves as a potential for the Eshelby stress tensor \cite{Eshelby1951,Kienzler1997} in the material configuration through a pull-back operation or, alternatively, in the form of the Eshelby energy-momentum tensor \cite{Eshelby1975} as a function of the direct deformation gradient \cite{Eshelby1999},
\begin{equation}\label{eq:Eshelby_stress_F}
\boldsymbol \Sigma\left(\mathbf{f}\right)=
\frac{\partial \psi\left(\mathbf{f}\right)}{\partial \mathbf{f}} \cdot \left[\text{cof} \, \mathbf{f}\right]^{-1}
=\Psi \mathbf{I}-\mathbf{F}^\text{T}\cdot \frac{\partial \Psi}{\partial\mathbf{F}},
\end{equation}
\noindent with $\text{cof} \, \mathbf{f} = \left[\det \mathbf{f}\right]\mathbf{f}^{-\text{T}}$ the cofactor of the reverse deformation gradient.

The energy-momentum representation in Equation~\ref{eq:Eshelby_stress_F} allows to calculate the Eshelby stress via post-processing based on the solution of the spatial motion problem. 

The material motion problem can be discretized via FE to obtain configurational forces at the nodes of the discretization by elements in ${\mathcal{B}_\text{e}} \in \Omega_0$ \cite{Steinmann2001}. The configurational force at global node $A$ can be calculated using the material gradient of the global node $A$ basis function ($N^A$), the Eshelby stress tensor ($\boldsymbol \Sigma$), and the FE assembly operator $\opA_{e=1}^{n_\text{el}}$ over all elements
\begin{equation}\label{eq:CForce_node}
\mathbf{F}^A_\text{CNF} = \opA_{e=1}^{n_\text{el}} \int_{\mathcal{B}_\text{e}} \boldsymbol \Sigma \cdot \nabla_0 N^A \text{d}\text{V}.
\end{equation}

\subsection{Analysis of configurational forces to assess mixed-mode fracture}

\begin{figure}[htbp]
\centering
	\includegraphics[width=0.95\columnwidth]{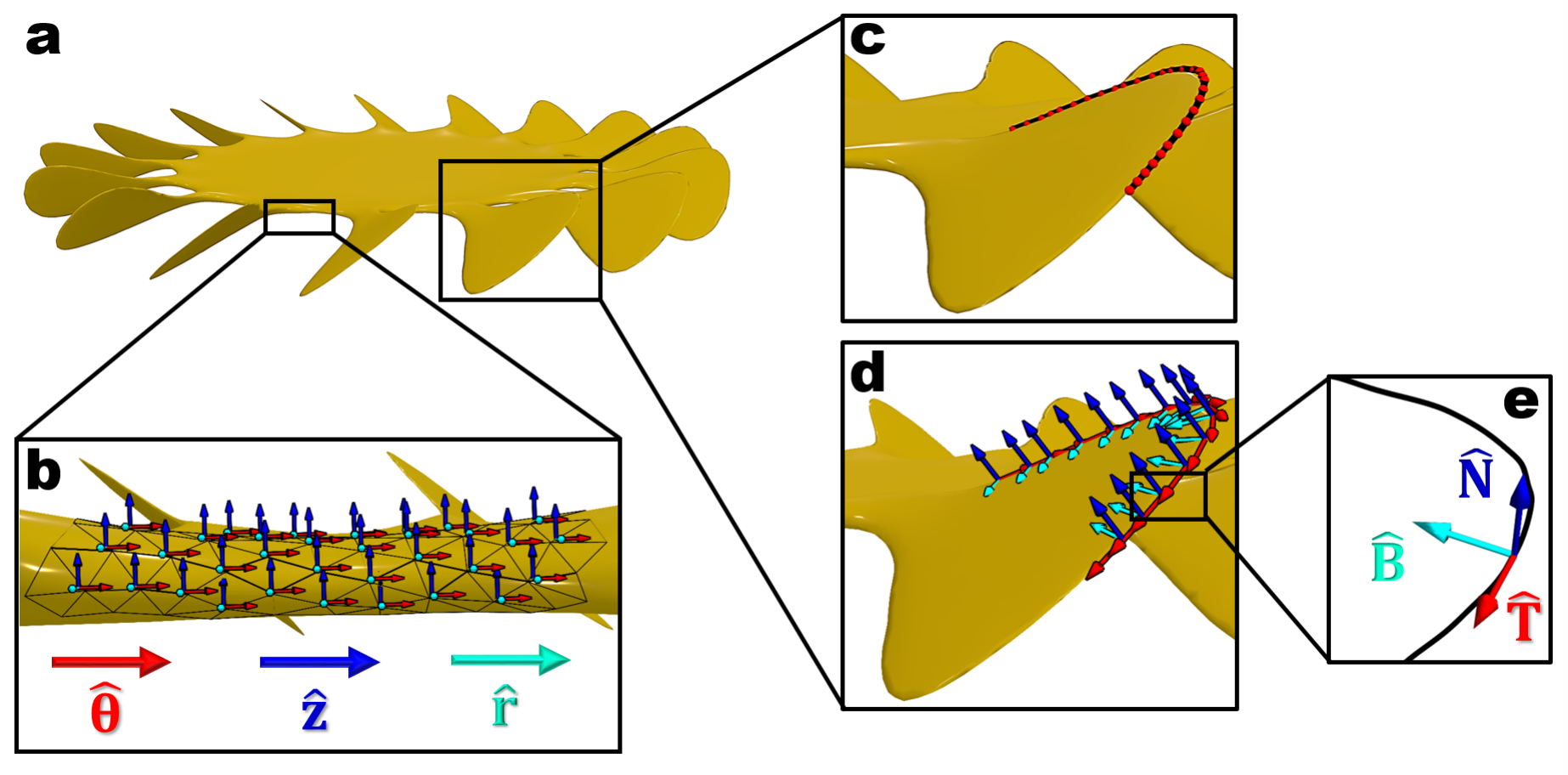}
    \caption{\label{fig:changeBase} \textbf{Configurational forces analysis.} (a) Idealized geometrical model of a mixed-mode fracture for $\phi=40\degree$. The rectangles highlight the regions of the crack front where configurational forces are computed: facets and planar segments. (b) Local coordinate frame at points along the crack front within a planar segment. The radial vector is shown for visualization purposes; in practice, the configurational force vector points into the material void. (c) Nodes along the crack edge on a facet where configurational forces are calculated. The black line represents the arc defined by these nodes. (d) Local coordinate frame at a facet tip used to resolve the force vector. (e) Close-up view of a single facet, where the crack front is represented as a black 3D curve, with the local coordinate frame depicted at a specific point along the front. A reduced number of points is shown in panels (b)–(d) for clarity.}
\end{figure}

We focus on computing the configurational forces along the crack front, i.e., in two regions: at points on the planar segments — the regions between facets belonging to the parent crack — and at points on the edges of the facets (Fig.~\ref{fig:changeBase}a).
We determine the total configurational force, $\mathbf{F}_\text{CNF}$, per unit length, $s$, by summing the forces acting on the nodes along the crack front and dividing by the arc length defined by these nodes, as illustrated in Fig.~\ref{fig:changeBase}c. In addition to the physical configurational forces at the crack front edge, we also account for configurational forces in a region surrounding the crack tip. These additional forces—sometimes referred to as spurious configurational forces—arise due to finite element discretization and lack direct physical meaning in the continuum sense. However, their inclusion enhances the approximation of the total configurational force, as described in \cite{Denzer2003}.

At each node, the components of the configurational force are computed in a local coordinate system defined at the crack front. For planar segments—i.e., the regions between facets belonging to the parent crack—the configurational force vector is transformed into cylindrical coordinates. This transformation uses an orthonormal basis formed by unit vectors in the radial (\(\bm{\hat{r}}\)), axial (\(\bm{\hat{z}}\)), and azimuthal (\(\bm{\hat{\theta}}\)) directions (Fig.~\ref{fig:changeBase}b). For the facets, the configurational forces are decomposed in the \((\bm{\hat{T}}, \bm{\hat{N}}, \bm{\hat{B}})\) orthonormal basis. It consists of unit vectors: \(\bm{\hat{T}}\) tangential to the crack front, \(\bm{\hat{N}}\) normal to the crack front, and \(\bm{\hat{B}}\) in the binormal direction, forming a right-handed coordinate system (Fig.~\ref{fig:changeBase}d and e).

\subsection{Numerical implementation}

The weak form of Equation~\ref{eq:strong_form} is solved with the open-access FE suite FEniCS \cite{Logg2012} and the solution is visualized with Paraview. The volumetric mesh of the sample is generated by tetrahedralization of the material's occupied volume. The idealized crack model represents the internal fracture surface within the simulation domain, while a cylindrical surface defines its outer boundaries. The mesh is fine at the crack front vicinity to render an accurate discretization of the Eshelby stress. The displacement field is discretized using a linear approximation. 

In all cases, the imposed Dirichlet boundary conditions simulate those from reported experiments \cite{ortellado2025principle}. The upper and lower bases are displaced vertically in opposite directions, while out-of-plane shear is introduced by rotating the upper base while keeping the lower base fixed. First, tensile stress is applied, with displacement $\Delta L$ increasing steadily from $0 \, \text{mm}$ to $4 \, \text{mm}$. In the following, this is referred to as mode-I phase. Subsequently, torsion is applied in incremental steps, with the torsion angle $\alpha$ gradually increasing from $0\degree$  to $90 \degree$. This will be denoted as mode I+III phase. For the aforementioned virtual experiment, we compute the resulting configurational force per unit length along the crack front, $\mathbf{F}_{\text{CNF}}/s$, as a function of $\Delta L$ and $\alpha$. Mesh resolution and neighborhood size were selected, ensuring that computed quantities remained stable across variations.

\section{Configurational forces reveal the role of facet tilting and local shear in determining propagation direction}

\subsection{Results for configurational forces on facets with varying tilting angles under mixed-mode I+III loading}

\begin{figure}[htbp]
 \centering
 \includegraphics[width=0.82\textwidth, keepaspectratio]{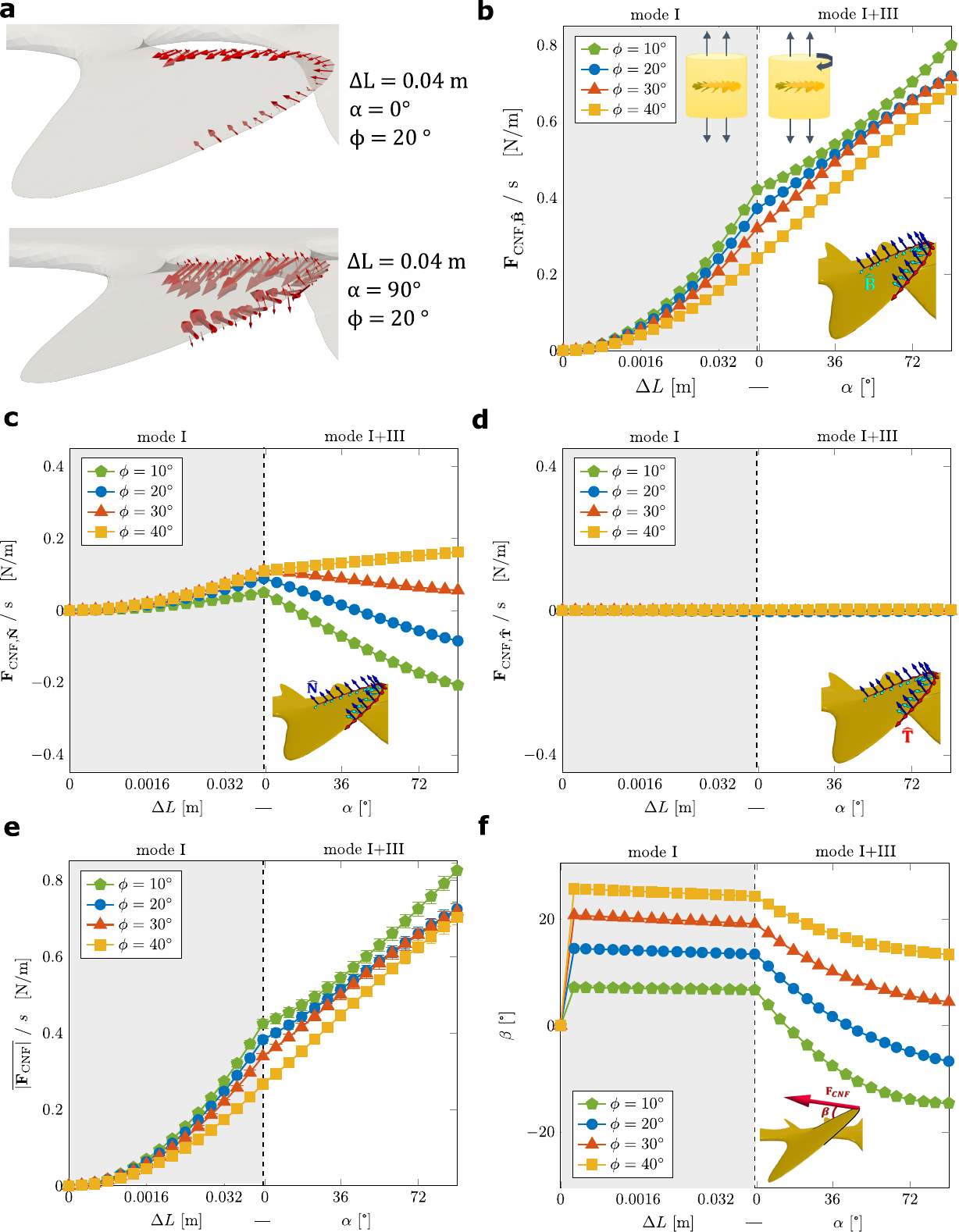}
 \caption{\textbf{Facet tilting angle $\phi$ controls the direction and strength of the configurational force under mixed-mode loading.} 
 The configurational force (CF) is computed by summing individual nodal forces along the facet tip and dividing by the length of the crack tip contour, including spurious forces in the vicinity. Due to the adopted variational formulation, the CF is oriented opposite to the direction of crack propagation. For practical purposes, the most favorable growth direction requires permuting the direction of the CF components. 
 (a) Two snapshots showing CF vectors at the facet tip: at maximum uniaxial tension (top) and at combined maximum tension and torsion (bottom).
 (b–d) Evolution of the CF components in the binormal \(\bm{\hat{B}}\), normal \(\bm{\hat{N}}\), and tangent \(\bm{\hat{T}}\) directions, respectively. (e) CF modulus. 
 (f) Angle $\beta$ between the CF vector and the facet plane, calculated along the loading of the cylinder in (f), better illustrates the relative direction of the force. 
CF values are averaged across all facets, and error bars represent the standard deviation of the mean. Schematic insets illustrate the applied loading, CF components in the local coordinate system and the angle $\beta$.}
 \label{fig:CF_facets_different-tilting}
 \end{figure}

We start by studying the evolution of the normalized configurational force magnitude, $\mathbf{F}_{\text{CNF}}/s$, for segmented facets, along the virtual experiment: mode I (tensile deformation) and mode I+III (tension + torsion) phases (Fig.~\ref{fig:CF_facets_different-tilting}). The force vector components in the local normal \(\bm{\hat{N}}\), tangent \(\bm{\hat{T}}\), and binormal \(\bm{\hat{B}}\) directions, as well as the vector magnitude, are presented. For all values of the tilting angle, $\phi$, the force magnitude, $|\mathbf{F}_{\text{CNF}}|$, increases with applied tension and is further amplified by the superposition of torsion. This trend is particularly evident in the simulation snapshots (Fig.~\ref{fig:CF_facets_different-tilting}a). $|\mathbf{F}_{\text{CNF}}|$ is consistently higher for lower values of $\phi$ under both uniaxial tension and combined tension-torsion loading, with the difference between curves decreasing as torsion is applied (Fig~\ref{fig:CF_facets_different-tilting}b). The closer a facet is to a planar orientation, the more effectively it aligns with the Mode I opening stress, thereby maximizing the configurational force magnitude. This behavior is consistent with the crack front propagating along orientations where the local energy release rate is highest — occurring when the facet orientation optimally captures the applied Mode I opening stress.

The configurational force's magnitude quantifies the cracks' propensity to advance, while its orientation indicates the preferred propagation direction. It is important to note that, in this work, the configurational force is oriented opposite to the direction of crack growth due to the adopted variational principle. For practical purposes, the most favorable direction of crack front propagation requires permuting the direction of the configurational force components. $\mathbf{F}_{\text{CNF},\mathbf{\hat{B}}}$ is the dominant component, whereas $\mathbf{F}_{\text{CNF},\mathbf{\hat{T}}}$ remains negligible. This indicates that the facets exhibit a propensity to propagate maintaining their tilting angle. Interestingly, $\mathbf{F}_{\text{CNF},\mathbf{\hat{N}}}$ is non-zero. For all $\phi$, $\mathbf{F}_{\text{CNF},\mathbf{\hat{N}}}$ increases under pure tension and decreases for most cases when torsion is applied. To better understand the effect of a non-vanishing normal component $\mathbf{F}_{\text{CNF},\mathbf{\hat{N}}}$, we compute the angle $\beta$ between $\mathbf{F}_{\text{CNF}}$ and the facet plane (Fig~\ref{fig:CF_facets_different-tilting}f). Under pure tension, $\beta$ increases for all $\phi$---under the restriction that $\beta \leq \phi$ since the crack front may not propagate downwards---, and decreases under added torsion. Notably, for $\phi \leq 20^\circ$ and large torsion angles, $\beta$ reverses its direction. This suggests that $\mathbf{F}_{\text{CNF}}$ tends to align with the planar parent crack plane under pure tension, while it rotates towards the plane perpendicular to the planar crack when torsion is superimposed, i.e., negative value of $\mathbf{F}_{\text{CNF},\mathbf{\hat{N}}}$. 

Physically, a non-zero normal component, $\mathbf{F}_{\text{CNF},\mathbf{\hat{N}}}$, hints at a potential geometric instability in the crack propagation path, which may promote small deflections from the facet’s initial orientation and induce slight rotations in the facet propagation direction, even in the absence of in-plane shear. Qualitative fluctuations in the facet orientation can be observed in earlier experiments in hydrogels \cite{Santarossa2023, ortellado2025principle} and are visible in Fig~\ref{fig:example_mixed_mode}. A similar behavior can also be observed in fractures of PMMA samples in the work of Hattali et al. \cite{Hattali2021}. In experiments, hydrogels' cracks are typically observed to advance for torsion angles smaller than $90 \degree$ \cite{ortellado2025principle}. 

In the following sections, we will show that linear elastic fracture mechanics cannot fully account for the observed behavior, and that deviations from its predictions arise due local out-of-plane shear on the facet surface.

\subsection{LEFM-based scaling estimate of propagation orientation under mixed-mode loading} \label{Scaling}

To rationalize the trends observed in the orientation of the configurational force vector relative to the facet plane, we develop a scaling estimate grounded in classical linear elasticity. The aim is to estimate the angle $\beta$ as a function of the facet inclination angle $\phi$.\footnote{We emphasize that this scaling is not intended to provide quantitative predictions of $\beta$, but rather to qualitatively capture its trends with respect to loading mode and facet inclination.}

As a first-order approximation, we consider a single semi-circular tilted crack embedded in a homogeneous, infinite elastic medium. The crack is inclined at an angle $\phi$ with respect to the plane perpendicular to the loading direction (horizontal plane). The sample is first subjected to remote uniform tensile loading (Mode I), and torsion (Mode III) is subsequently applied after the tensile deformation is established. This setup approximates the energetic environment of an isolated facet, i.e., in the absence of interaction effects. Finite size effects and material non-linearities are not considered.  

Under pure Mode I loading, the crack experiences opening stresses normal to the global fracture plane. When projected onto an inclined facet, the local stress intensity factor is reduced due to geometric misalignment. For the considered crack under uniaxial tension, the Mode I intensity scales as ~\cite{murakami1985}:
\begin{equation}
    K_I(\phi) \sim K_{I,0} \cos^2 \phi,
\end{equation}
where $K_{I,0}$ denotes the Mode I stress intensity factor for a planar (i.e., $\phi = 0$) crack.

In the absence of torsion, the inclined crack also experiences an in-plane shear component (mode~II) owing purely to its geometric misalignment, according to:

\begin{equation}
    K_{II}(\phi) \sim   K_{I,0}\sin\phi\cos\phi. 
\end{equation}

In the presence of torsion, a Mode III component arises due to the remote antiplane shear field. For a tilted facet, the amplitude of this shear mode scales geometrically with inclination as ~\cite{Molnar2024}:
\[
K_{III}(\phi) \sim K_{III,0} \sin2\phi,
\]
where \( K_{III,0} \) is a nominal amplitude characterizing the far-field torsional loading.

The configurational force magnitude is proportional to the energy release rate, which in linear elasticity is given by:
\[
G \sim \eta_I [K_I^2(\phi) +K_{II}^2(\phi)]+ \eta_{III} K_{III}^2(\phi),
\]
with \( \eta_I = \frac{1 - \nu^2}{E} \) and \( \eta_{III} = \frac{1 + \nu}{2\mu} \). These expressions represent the elastic prefactors for Modes I and III, respectively. Accordingly, the local configurational force magnitude scales as:
\[
|\mathbf{F}_{\text{CNF}}| \sim \eta_I [K_I^2(\phi) +K_{II}^2(\phi)] + \eta_{III} K_{III}^2(\phi).
\]

Decomposing the force vector into components in the local coordinate system of the crack front yields:
\begin{align}
    \mathbf{F}_{\text{CNF},\mathbf{\hat{N}}} &\sim \eta_I \left[ K_I^2(\phi) + K_{II}^2(\phi) \right] \sin \phi, \\
    \mathbf{F}_{\text{CNF},\mathbf{\hat{B}}} &\sim \eta_I \left[ K_I^2(\phi) + K_{II}^2(\phi) \right] \cos \phi + \eta_{III} K_{III}^2(\phi).
\end{align}

Defining the loading ratio \( \omega = K_{III,0} / K_{I,0} \) and the prefactor ratio \( \gamma = \eta_{III} / \eta_I = \frac{[1 + \nu]^2}{1 - \nu^2} \), we obtain:
\[
\tan \beta = \frac{\mathbf{F}_{\text{CNF},\mathbf{\hat{N}}}}{\mathbf{F}_{\text{CNF},\mathbf{\hat{B}}}} \sim \frac{\cos^2 \phi \sin \phi}{\cos^3 \phi + \gamma \omega^2 \sin^22\phi}.
\]

Under pure tensile loading, the equation yields: 
\begin{equation}
    \tan \beta \sim \tan \phi \longrightarrow \beta \approx \phi.
\end{equation}

Despite its simplicity, the scaling model captures the trends observed in our simulations. Qualitatively, for a given $\phi$, the scaling predicts that $\beta$ remains constant under increasing uniaxial tension and decreases as torsion is added. This trend aligns with the behavior observed in our simulations (Fig.~\ref{fig:CF_facets_different-tilting}). However, the scaling predicts $\beta \approx \phi$ under pure tensile loading, while our simulations consistently yield $\beta < \phi$. Moreover, under mixed Mode I+III loading, the scaling predicts $\beta \geq 0$ for all values of the loading ratio $\omega$ and the coefficient $\gamma$, with $\phi \in [0^\circ, 90^\circ]$. In contrast, our simulations reveal that $\beta$ becomes negative for $\phi \leq 20^\circ$.

In summary, the scaling model serves as a simplified reference framework, capturing leading-order trends in $\beta$. In the next section, we explore deviations from this scaling by examining local shear deformation on the facets.

\subsection{Local shear causes deviations from linear elasticity predictions}

In the absence of curvature or twisting, a tilted facet embedded in an infinite linear elastic solid and subjected to remote tensile loading would be expected to carry only tensile and in-plane shear stresses. However, our simulations consistently show that the direction of the configurational force vector deviates from this ideal behavior, even under nominally pure Mode I loading. This suggests that additional local stress components may be present.

\begin{figure}[htbp]
 \centering
 \includegraphics[width=1.0\textwidth, keepaspectratio]{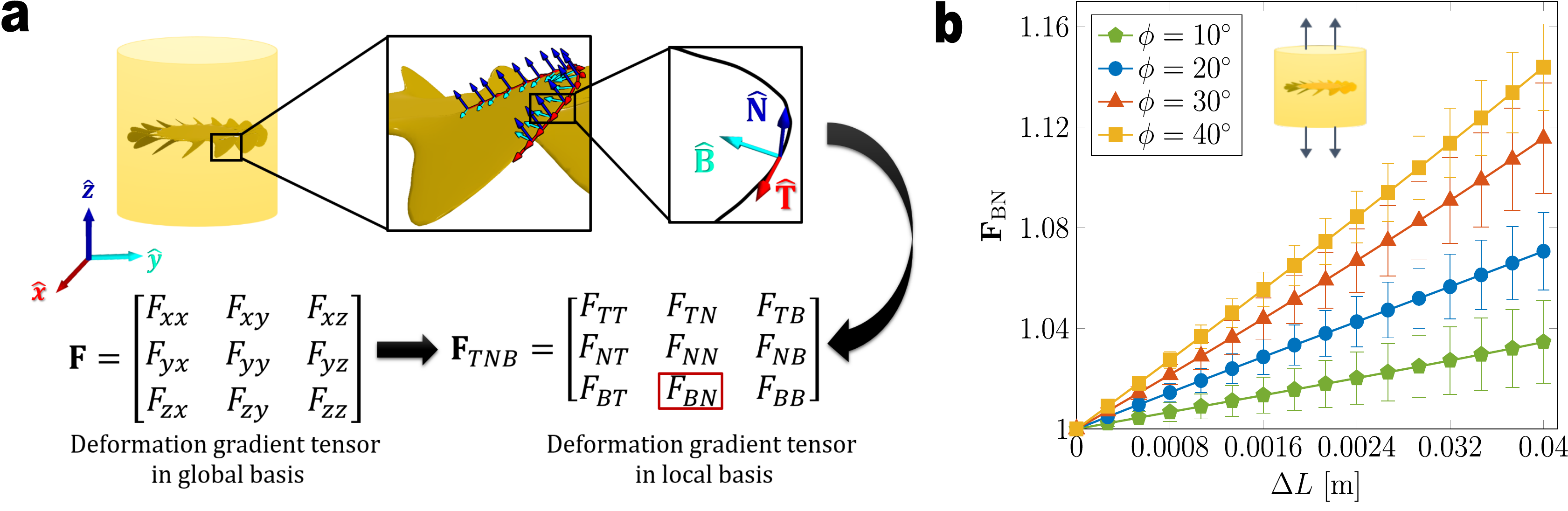}
 \caption{\textbf{Shear deformation in the binormal direction increases with facet tilting angle under applied Mode I tension.} 
 a) Schematic of the local crack-front basis (\( \bm{\hat{T}}, \bm{\hat{N}}, \bm{\hat{B}} \)) in which the deformation gradient tensor is computed. 
 b) Evolution of the deformation gradient component $\mathbf{F}_{BN}$ computed in the local crack-front basis, under increasing tensile displacement for cracks with different facet tilting angles \( \phi \). The component $\mathbf{F}_{BN}$ quantifies the out-of-plane shear induced by deformation along the facet normal direction. Its value is consistently non-zero and increases both with the applied Mode I load and with facet inclination. This local shear may contribute to deviations in the configurational force orientation angle $\beta$ relative to the LEFM-based scaling predictions.
 At each loading step, the maximum value $\mathbf{F}_{BN}$ per facet is extracted. Plotted values show the mean across all the facets in a crack; error bars represent the standard deviation.
 }
 \label{fig:deformation_gradient}
 \end{figure}

To better understand these deviations from the LEFM-based scaling predictions, we compute the deformation gradient tensor $\mathbf{F}$, expressed in the orthonormal local frame defined by the tangent (\(\bm{\hat{T}}\)), normal (\(\bm{\hat{N}}\)), and binormal (\(\bm{\hat{B}}\)) directions (Fig.~\ref{fig:deformation_gradient}a). We focus on the off-diagonal component ($\mathbf{F}_{BN}$), which describes shear deformation in the binormal direction induced by gradients along the normal.

Figure~\ref{fig:deformation_gradient} shows the evolution of the maximum value of the deformation gradient component $\mathbf{F}_{BN}$ under increasing pure tensile deformation (Mode I phase) for cracks with varying facet tilting angles $\phi$. Since $\mathbf{F}_{BN}$ varies along the crack front, we compute the maximum value per facet at each loading step to quantify the most significant local shear deformation. The component $\mathbf{F}_{BN}$ is consistently non-zero across the loading path and increases with both applied tension and tilting angle. These results indicate the presence of localized shear deformation on the facet surface in the binormal direction. This effect is not captured by the LEFM-based scaling and arises from the geometric misalignment between the tilted facet and the loading direction. The resulting local stress gradients---despite nominally pure Mode I loading---may contribute to a reduction in the angle $\beta$, offering a mechanistic explanation for the observed deviations from scaling predictions.

We note that the proposed LEFM-based scaling considers only singular stress contributions and neglects non-singular terms such as the T-stress. While often treated as secondary, since they do not contribute to the energy release rate in LEFM, these non-singular components may still alter the stress distribution locally and, therefore, affect the stability of the crack and the preferred growth direction ~\cite{cotterell1980}.

In principle, differences between our simulation results and the LEFM-based scaling prediction (Section~\ref{Scaling}) may also arise from (i) elastic interactions between adjacent facets or (ii) the finite size of the gelatin cylinder. As shown in Section~\ref{Facet-spacing-beta}, both effects have only a limited influence on $\beta$ during the mode I phase and at low to intermediate torsion angles. Finite-size effects become non-negligible for large torsion angles ($\alpha > 36^\circ$), as discussed in Appendix~\ref{Facet-spacing-beta}.

\section{Configurational forces reveal how facet–planar region interactions govern type B crack growth under mixed-mode loading}

\subsection{Results for configurational forces on planar regions for cracks with different tilting angles under mixed-mode loading}

\begin{figure}[htbp]
\centering
\includegraphics[width=0.9\textwidth, keepaspectratio]{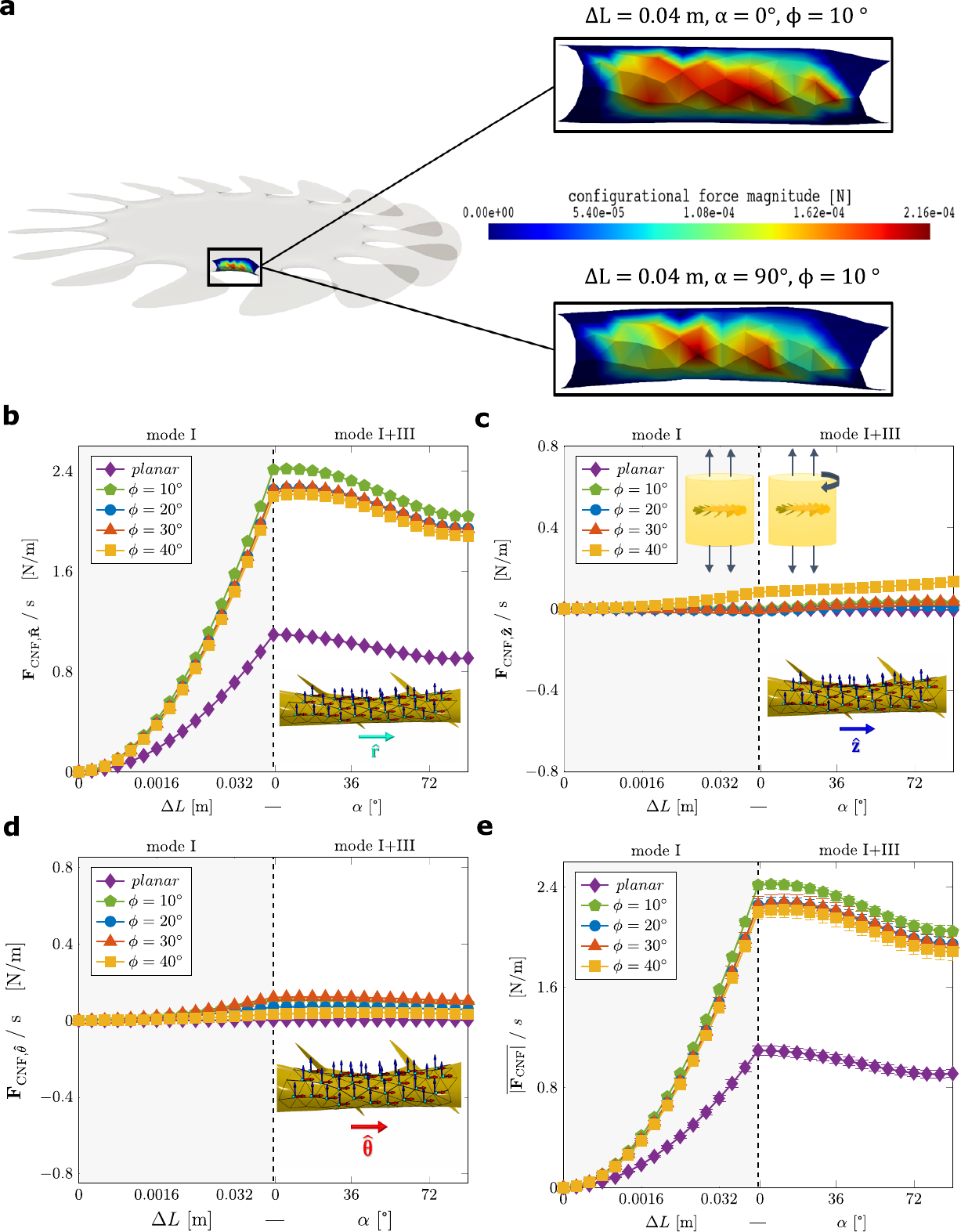}
\caption{\textbf{Superimposed torsion reduces the configurational force on planar regions, lowering their growth tendency under mixed-mode loading across different facet tilting angles.} 
(a) Crack mesh with planar segments color-coded by the magnitude of the configurational force, shown under pure tension (top) and combined tension with superimposed torsion (bottom). 
(b-d) Components of the configurational force vector in the local radial \(\bm{\hat{r}}\), axial \(\bm{\hat{z}}\), and tangent \(\bm{\hat{\theta}}\) directions, respectively. 
(e) Force modulus.
Results are shown for cracks with different facet tilting angles $\phi$. Configurational forces are averaged across all planar segments; error bars represent the standard deviation of the mean to quantify numerical scatter. Schematic insets depict the force components and applied loading configuration.}
\label{fig:CF_planar_different-tilting}
\end{figure}

We begin by analyzing the evolution of the normalized configurational force magnitude, $\mathbf{F}_{\text{CNF}}/s$, on planar segments—i.e., the regions of the parent crack between adjacent facets—under different loading conditions (Fig.~\ref{fig:CF_planar_different-tilting}). The force vector components in the local radial \(\bm{\hat{r}}\), axial \(\bm{\hat{z}}\), and tangential \(\bm{\hat{\theta}}\) directions, as well as the vector magnitude $|\mathbf{F}_{\text{CNF}}|$, are shown. Along the mode I tensile phase, it is observed that $\mathbf{F}_{\text{CNF},\mathbf{\hat{r}}}$ is dominant, while $\mathbf{F}_{\text{CNF},\mathbf{\hat{z}}}$ and $\mathbf{F}_{\text{CNF},\mathbf{\hat{\theta}}}$ are considerably smaller and close to zero. This indicates the tendency of the crack to grow in the radial direction. The legend ``planar''  refers to a penny-shaped fracture with the same radius and thickness as the parent crack, without facets ($\phi=0$). As expected, the force magnitude increases with tensile loading. However, in the mode I+III phase, it decreases for all nonzero values of $\phi$. This trend becomes clearly apparent in the configurational force maps in Fig.~\ref{fig:CF_planar_different-tilting}a, where $|\mathbf{F}_{\text{CNF}}|$ considerably decreases with the application of torsion. These results suggest that echelon cracks reduce the propensity for crack growth at the planar crack front under mixed-mode I+III loading, in line with experimental observations from Hattali et al. \cite{Hattali2021}, who reported increased fracture resistance associated with crack front segmentation in brittle materials under mixed-mode loading.

Interestingly, the magnitude of $\mathbf{F}_{\text{CNF}}/s$ for planar segments exceeds that of the facets across all loading conditions (maximum values are approximately four times higher). This suggests that the planar regions may become energetically favorable to propagate, potentially bridging the space between facets. In the studied crack model, facets are well-developed and far from the onset of facet nucleation. These results are consistent with experimental observations of Pham and Ravi-Chandar \cite{Pham2016}, who reported that initially, type A cracks (the facets) nucleate. Then, as they propagate, the stress intensity factor at the parent crack increases, leading to the growth of the bridging regions (type B cracks) and causing the parent crack front to advance toward the facets. Similar observations have been reported experimentally by Lin et al. \cite{Lin2010}, Santarossa et al. \cite{Santarossa2023} and Ortellado \cite{ortellado2025principle}. Notably, the configurational force on the planar segments not only exceeds that on the facets, but also surpasses the value observed for an unsegmented, planar, penny-shaped crack. In the following section, we investigate the origin of this amplification and examine how elastic interactions between facets and planar regions influence the growth tendency of the latter.

\subsection{Facet spacing modulates the growth of type B cracks}

We hypothesize that the differences in $|\mathbf{F}_{\text{CNF}}|/s$ between facets and planar segments arise from an amplification effect due to elastic interactions between them. To test this hypothesis, we use numerical simulations to compute $|\mathbf{F}_{\text{CNF}}|/s$ for cracks with a fixed facet tilt angle ($\phi = 20^\circ$) and varying facet spacing $\Lambda$. 

In our echelon crack model, facets are distributed equidistantly and symmetrically along the crack front, allowing the spacing between them to be controlled by adjusting the number of facets. The angular separation between adjacent facets $\Lambda$ is therefore defined as  \( \Lambda = 2\pi/n \), where \( n \) is the total number of facets in the fracture. The ‘planar’ case represents the limit of a facet-free, penny-shaped crack with the same size as the parent fracture, corresponding to infinite facet spacing ($\Lambda \to \infty$). 

\begin{figure}[htbp]
\centering
\includegraphics[width=0.6\textwidth, keepaspectratio]{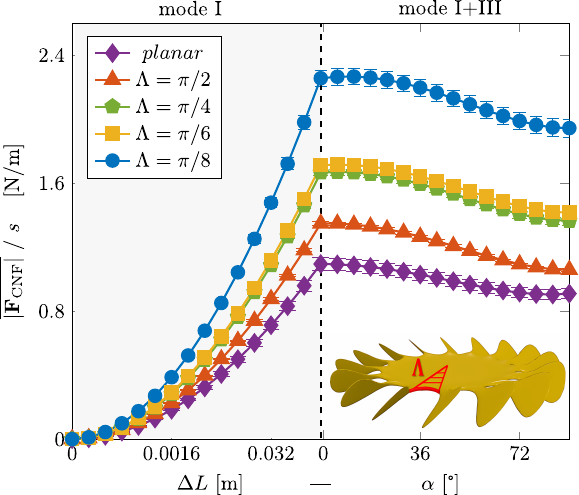}
\caption{\textbf{Facet spacing modulates the configurational force on planar segments.} 
Normalized configurational force magnitude $|\mathbf{F}_{\text{CNF}}|/s$ on planar crack segments is plotted as a function of facet spacing $\Lambda$, for cracks with fixed tilting angle $\phi = 20^\circ$. The legend ``planar''  refers to a penny-shape fracture with the same radius and thickness as the parent crack, without facets ($\Lambda \to \infty$).
As facet spacing increases, $|\mathbf{F}_{\text{CNF}}|/s$ decreases, indicating a reduction in elastic interactions between facets and planar regions.
This supports the hypothesis that closer facet proximity amplify the configurational driving force on planar segments, promoting type B crack growth.
Force values are averaged across all planar segments; error bars represent the standard deviation of the mean.
The schematic inset illustrates the angular facet spacing $\Lambda$.
}
 \label{fig:CF_planar_spacing}
\end{figure}

As shown in Figure ~\ref{fig:CF_planar_spacing}, $|\mathbf{F}_{\text{CNF}}|/s$ is consistently higher on planar segments for cracks with smaller facet spacing, across the full loading path. This trend indicates that when facets are closely spaced, the resulting elastic interactions amplify the configurational forces on the planar regions. In contrast, as spacing increases, these interactions weaken, reducing the configurational force of planar segments. Together, these results suggest that facet–facet spacing modulates the amplification of configurational forces on planar regions, thereby influencing the emergence and propagation of type B cracks (planar segments).

\subsection{Type B crack nucleation induces asymmetric configurational force redistribution}

\begin{figure}[htbp]
 \centering
 \includegraphics[width=1.0\textwidth, keepaspectratio]{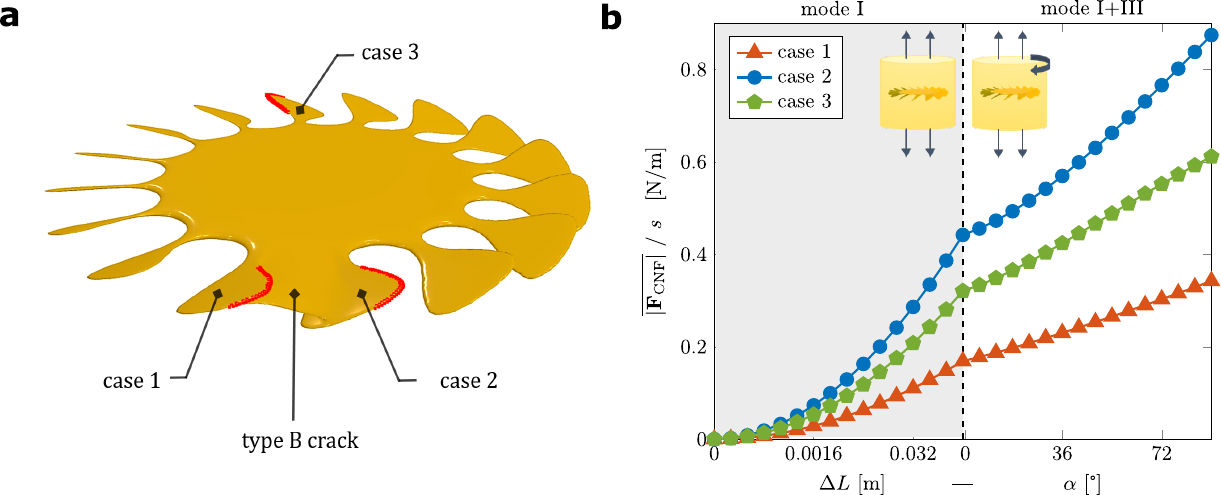}
 \caption{\textbf{Asymmetric redistribution of configurational forces due to facet coalescence.} 
 (a) Schematic of a crack with a nucleated type B crack. Red points indicate mesh nodes where configurational forces are computed. This mesh model was generated using a molecular dynamics-based mesh relaxation method to ensure a smooth union of the facets and the type B crack, allowing accurate evaluation of configurational forces (see ~\ref{Molecular-dynamics}). The three analyzed regions are highlighted: (1) near the type B crack, (2) farther from the coalescence zone on the adjacent facet, and (3) on a reference facet away from the coalesced pair.
 (b) Normalized configurational force magnitude, $|\mathbf{F}_{\text{CNF}}|/s$, for the three cases. The comparison reveals shielding near the type B crack and amplification farther away, indicating a nonlocal redistribution of driving forces induced by coalescence.} 
 \label{fig:CF_facet_coalescense}
 \end{figure}

In addition to the influence of facet spacing, we investigate how the nucleation of a type B crack affects the configurational forces acting on adjacent facets. To this end, we analyze the evolution of the normalized configurational force magnitude, $|\mathbf{F}_{\text{CNF}}|/s$, for selected facets throughout the virtual experiment—including both the Mode I (tensile deformation) and Mode I+III (tension + torsion) phases (Fig.~\ref{fig:CF_facet_coalescense}). We focus on cracks with a facet tilting angle $\phi = 20^\circ$. Three scenarios are examined, as illustrated in Figure ~\ref{fig:CF_facet_coalescense}a: (1) a facet region near the coalescence zone (type B crack), (2) a region on the other coalescing facet but farther from the type B crack, and (3) a facet located away from the coalesced pair, which serves as a reference for comparison.

Compared to the reference case, the facet region near the type B crack (case 1) exhibits a notable reduction in the configurational force magnitude, indicating a shielding effect. In contrast, the region far from the coalescence zone on the adjacent facet (case 2) experiences force amplification, showing a significant increase in $|\mathbf{F}_{\text{CNF}}|/s$.

Prior to coalescence, adjacent facets exhibit similar configurational force magnitudes. After coalescence, this balance is disrupted: depending on proximity to the new segment, facets show either amplification or shielding of the configurational force. This asymmetry reflects a local breaking of force distribution symmetry induced by the geometric reconfiguration.

Overall, these results suggest that the presence of a type B crack has a nonlocal influence: it shields nearby facet regions while amplifying the driving forces in more distant regions of the coalescing facets.

\section{Facet spacing regulates the strength of driving forces for facet propagation}

\begin{figure}[htbp]
 \centering
 \includegraphics[width=1.0\textwidth, keepaspectratio]{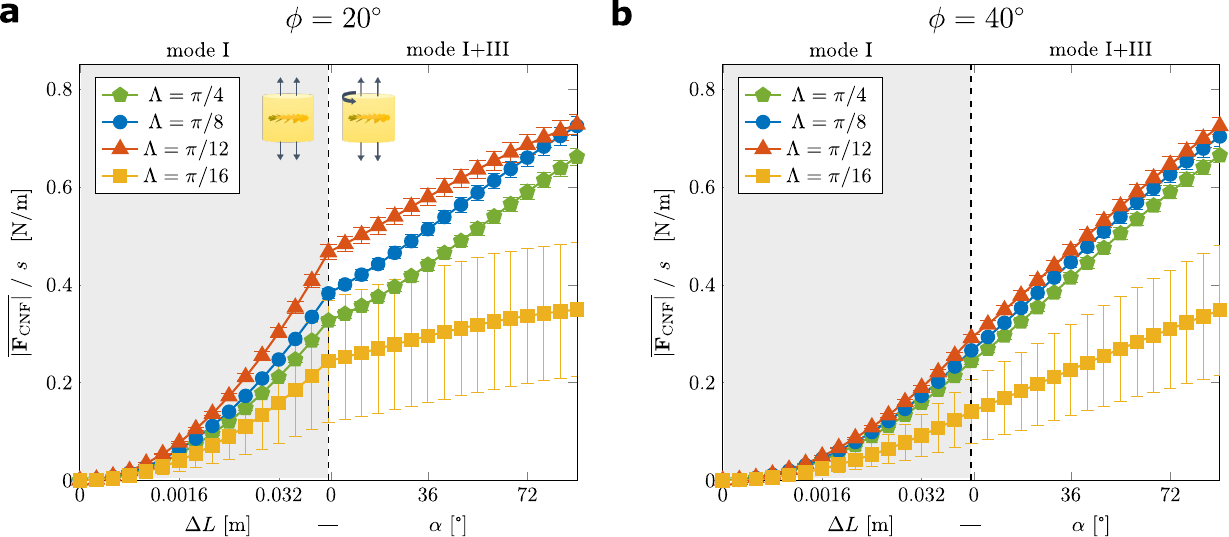}
 \caption{\textbf{Facet spacing modulates the configurational force at facet edges.}  Normalized configurational force magnitude $|\mathbf{F}_{\text{CNF}}|/s$ on facets as a function of facet spacing $\Lambda$, for cracks with fixed tilting angles: (a) $\phi = 20^\circ$ and (b) $\phi = 40^\circ$.
As facet spacing decreases, $|\mathbf{F}_{\text{CNF}}|/s$ increases along the loading path. When the spacing falls below a critical value $\Lambda_c$, the force magnitude is significantly reduced. This behavior suggests that elastic interactions between neighboring facets amplify the local configurational force up to a threshold spacing, below which shielding effects dominate and reduce the force intensity.
At small facet spacing, mesh asymmetries lead to increased numerical scatter. However, the trend in $|\mathbf{F}_{\text{CNF}}|/s$ with $\Lambda$ remains identifiable and consistent across loading steps.
CF values are averaged over all facets, with error bars representing the standard deviation of the mean to quantify numerical scatter. Schematic insets illustrate the applied loading conditions.}
 \label{fig:CF_facets_different-spacing}
 \end{figure}

In this section, we study how the spacing between facets influences the configurational force acting on the edges of the facets. To this end, we analyze the evolution of the normalized configurational force magnitude, $|\mathbf{F}_{\text{CNF}}|/s$, for the facets throughout the virtual experiment—including both the Mode I (tensile deformation) and Mode I+III (tension + torsion) phases—for varying angular facet spacings $\Lambda$ (Fig.~\ref{fig:CF_facets_different-spacing}). We focus on two facet tilting angles: $\phi = 20^\circ$ and $\phi = 40^\circ$.

For both values of $\phi$, and across the entire loading path, $|\mathbf{F}_{\text{CNF}}|/s$ increases as facet spacing decreases. However, when the spacing falls below a critical threshold $\Lambda_c$, the force magnitude sharply decreases, independent of the tilting angle. 

This behavior can be interpreted as resulting from elastic interactions between neighboring facets. At intermediate spacings, closer proximity reinforces the interaction, leading to a constructive amplification of the configurational force. When facets approach overlapping configurations, however, these interactions become destructive, and shielding effects dominate, reducing the magnitude of the configurational force. 

These results align with recent findings \cite{ortellado2025principle}, which numerically show that facet–facet interactions in a simplified model of mixed-mode I+III cracks can either amplify or shield stress intensity factors along the crack front, depending on the tilting angle and spacing between facets. Furthermore, this amplification–shielding transition is consistent with the findings of Thomas et al. ~\cite{thomas2017quantification}, who showed, using FEM simulations of circular embedded cracks, that the stress intensity factors (SIFs) at a crack front are highly sensitive to the spatial arrangement of nearby fractures. When two crack tips are closely aligned (tip-to-tip), the local stress fields overlap constructively, leading to SIF amplification. In contrast, when cracks are oriented face-to-face (stacked), the interaction reduces the local driving force, resulting in shielding.

Overall, the results of this section suggest that facet spacing is not only a geometric parameter but a critical control variable governing the balance between facet amplification and shielding, and thus, the stability of segmented crack growth.

\section{Discussion and outlook}
This study presents a novel approach to crack segmentation in mixed-mode I+III fractures within the framework of configurational mechanics. The investigations in this work leverage finite-element discretized configurational forces to understand and explain the growth of echelon cracks---magnitude and direction---in a neo-Hookean gelatin hydrogel. This framework captures effects such as the reorientation of crack facets under mechanical loading and the resulting effects of facet coalescence—phenomena that cannot be explained using LEFM-based approaches. Our findings show that the reorientation of facets arises from the local shear deformation state at the crack front. 

In addition to facet orientation, our analysis reveals that the spacing between neighboring facets significantly influences the magnitude of the configurational forces. For intermediate spacings, elastic interactions between facets amplify the driving force at the crack front. However, when the spacing falls below a critical threshold, force magnitudes are reduced, indicating a shielding effect.  

Finally, we show that interactions between adjoining facets modulate—and, depending on facet spacing, can amplify—the configurational forces in the planar regions between them, thus promoting the nucleation of type B cracks. Building on this, we investigate the coalescence of two facets driven by the growth of a type B crack, and find that this process leads to shielding of the growth tendency in nearby facet regions, while amplifying the driving forces in more distant areas of the coalescing facets. 

Unlike conventional evaluations of fracture configurational forces via contour integration or domain evaluation method (see, e.g., \cite{Li1985}), the direct calculation of configurational forces at the nodes of the FE mesh enables studying fracture on geometrically intricate crack fronts and for virtually any material behavior. In this regard, a future direction is the direct application of the configurational force method to FE meshes reconstructed from experimental data----e.g., obtained via X-ray tomography---, enabling a seamless integration of modeling and measurement for the analysis of fracture and material instabilities under arbitrary mixed-mode conditions. Thus, configurational forces may be calculated directly on a number of defects and overall heterogeneities to assess the driving forces for fracture and even benchmark them to a fracture-toughness value.

Although crack propagation is not directly simulated in this work, the computed configurational forces can be interpreted in the context of fracture toughness and crack initiation. Specifically, the magnitude of the configurational force per unit length, \( |\mathbf{F}_\text{CNF}|/s \), quantifies the energetic driving force available for crack extension at each point along the crack front. Crack growth is expected to initiate when the configurational force exceeds a critical value, \( F_{\text{CNF,c}} \). This threshold is considered a material-inherent property, independent of geometry, stress state, or loading, as illustrated in \cite{Moreno-Mateos2024b} for mode-I fracture and in \cite{Moreno-Mateos2025} for fractures under biaxial loading conditions.

Our results indicate which regions of the crack front (facets vs. planar segments) are closer to satisfying this criticality condition. The increasing magnitude of configurational force per unit length on planar segments under either uniaxial tension or mixed-mode loading suggests that these regions reach the propagation threshold before the facets. This result is consistent with experimental observations of type B crack formation (cf. \cite{Lin2010, Pham2016}). 

Other future perspectives include the integration of the configurational force method and models for fracture. A diffusive crack, e.g., as modeled in phase-field models \cite{Molnar2024} may entail further theoretical developments and computational procedures to compute the effective configurational force (\cite{Bishara2024,Moreno-Mateos2024a}). A blunt crack modeled, for instance, with cohesive elements---or even a full discontinuous Galerkin FE discretization---may facilitate the computation of nodal configurational forces. In this case, however, the crack is forced to propagate across the FE elements' interfaces. X-FEM techniques or even peridynamics may provide an alternative framework for such implementations \cite{Steinmann2023}. The catalog is broad and the integration of the configurational force method and a model for fracture in a hybrid framework is promising.

\medskip

\section*{Data and Software Availability}
Data and software will be available upon reasonable request.

\section*{Acknowledgments}
Funded by the European Union. Views and opinions expressed are however those of the author(s) only and do not necessarily reflect those of the European Union or the European Research Council Executive Agency. Neither the European Union nor the granting authority can be held responsible for them. This work is supported by the European Research Council (ERC) under the Horizon Europe program (Grant-No. 101052785, project: SoftFrac). 
\begin{figure}[htbp]
\includegraphics[width=0.3\textwidth]{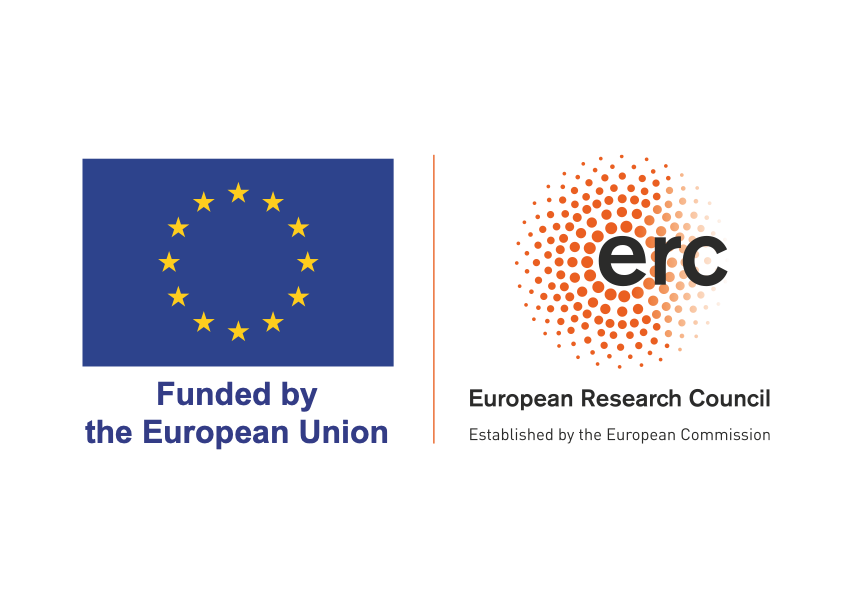}
\end{figure}

\section*{Competing Interests}
\noindent The Authors declare no Competing Financial or Non-Financial Interests.

\appendix
\section{Constitutive behavior of a benchmark hydrogel} \label{Model-calibration}

We consider a thermoreversible hydrogel composed of commercial bovine gelatin, water, and glycerin as a reference material for our research, following previous experimental studies on fracture under mixed loading \cite{Santarossa2023, ortellado2025principle}.

Gel samples were prepared for mechanical characterization to calibrate the numerical model. The gel was synthesized following the procedure described in \cite{Santarossa2023}, except that a Teflon mold was used for casting the gel instead of the container described in the original study.

\begin{figure}[htbp]
 \centering
 \includegraphics[width=0.9\textwidth]{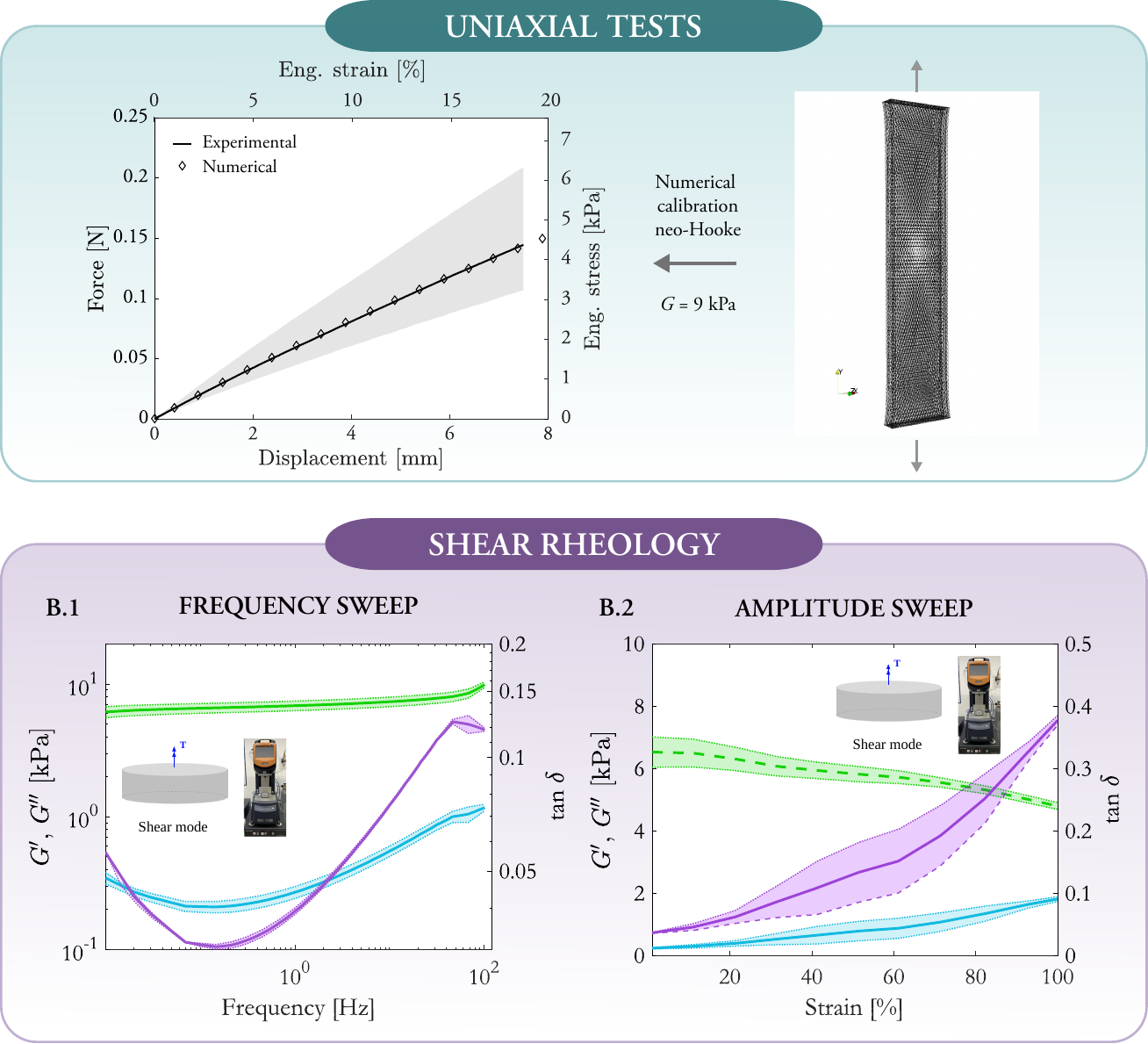}
 \caption{\textbf{Experimental characterization of the gelatin hydrogel.} (A) Uniaxial tensile tests. A FE simulation mimicking the experimental samples and loading conditions is used to calibrate the neo-Hookean constitutive model, which will be applied later in this work. (B) Rheological characterization of the gelatin. The shear storage ($G'$) and loss  moduli ($G''$) were measured. (B.1.) Frequency sweep tests for a constant strain of \qty{5}{\%}, from a frequency of \qty{0.01}{\hertz} to \qty{100}{\hertz}. (B.2.) Amplitude sweep tests for a constant frequency of \qty{0.0318}{\hertz}. All tests were performed at room temperature, i.e., \qty{25}{\celsius}.}
 \label{fig:experimental}
 \end{figure}

A universal tensile machine (Inspekt S 5 kN, Hegewald \& Peschke, Nossen, Germany) was used for mechanical testing. Rectangular samples were prepared with a width of \qty{11}{\milli \meter}, thickness \qty{3}{\milli \meter}, and initial length \qty{40}{\milli \meter}. Uniaxial loading was applied at a quasi-static loading rate of \qty{0.04}{\milli \meter \per \second}, corresponding to a strain rate of  \qty{0.001}{\per \second}, with force-displacement data recorded until sample rupture. The material exhibits linear-elastic behavior up to failure, with no observed yielding (Fig.~\ref{fig:experimental}A). To enhance the accuracy of the calibration of the mechanical parameters, 3D simulations were conducted to replicate the actual boundary conditions of the gel specimens, with the horizontal displacement at the upper and lower edges constrained to zero, reflecting the clamping in the experimental setup (Fig.~\ref{fig:experimental}A).

Rheological characterization of the gelatin was conducted using a Discovery HR-30 rheometer (TA Instruments, New Castle, DE, USA). Cylindrical samples for shear rheology measurements were prepared with a height of \qty{3500}{\nano \meter} and a diameter of \qty{40}{\milli \meter}. A slight axial pre-compression of \qty{1}{\newton} was applied to ensure proper contact between the upper geometry and the sample. The rheology results (Fig.~\ref{fig:experimental}B) show that the gelatin hydrogel exhibits solid-like behavior with a dominant elastic response. These observations, along with tensile test results, support the use of a Neo-Hookean constitutive model to simulate the gel’s mechanical behavior.

\section{Mesh generation of the simplified geometrical representation of echelon cracks} \label{Model-crack-mesh-description}

The crack mesh consists of a central planar region (parent crack), modeled as a thin, flat cylinder, and tilted segments (daughter cracks) extending from the parent crack. The facets are initially created separately and then integrated into the planar fracture.

\begin{figure}[htbp]
    \centering
    \includegraphics[width=1.0\textwidth, keepaspectratio]{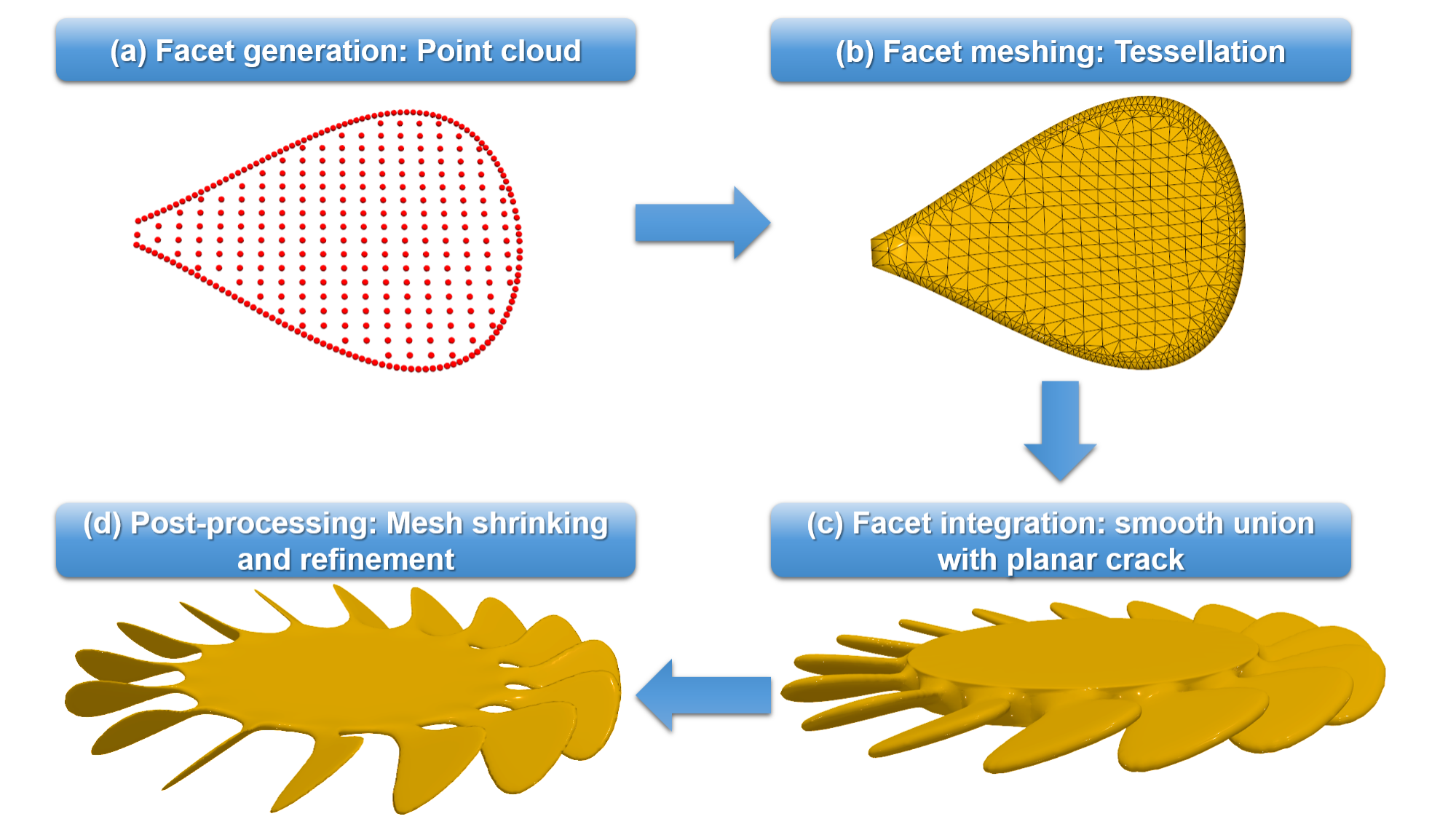}
    \caption{\textbf{Mesh generation pipeline for the simplified echelon crack geometry.} 
    (a) Point cloud defining the facet shape using a mathematical function. 
    (b) Triangulated facet mesh via Delaunay tessellation. 
    (c) Integration of tilted facets into a planar parent crack using smooth Signed Distance Field-based Boolean operation. 
    (d) Final mesh after shrinking and refinement steps.}
    \label{fig:crack_mesh_generation}
\end{figure}

The facet meshes are generated using the following procedure. First, the 2D profile is defined as a cloud of points (Fig.~\ref{fig:crack_mesh_generation}a). To achieve this, the contour of a facet is delineated using a mathematical function. In our case, we use:

\[
f(y) = \frac{\pm\sqrt{1 - 2y + 2y^3 - y^4}}{2}, \quad \text{for } y \in [-1, 0.8] ,
\]

to obtain a finger-like shape, as observed in experiments. Additional points are added along the vertical line \( x = 0.8 \) within the range \( y \in [-1,1] \) to create a closed region. Multiple layers of points are then stacked to form a 3D structure. The contours and enclosed regions of the outermost layers are populated with uniformly distributed points. The mesh of a facet is finally obtained using Delaunay tessellation (Fig.~\ref{fig:crack_mesh_generation}b).

Facets are placed along the edge of the parent planar crack at equal azimuthal distances. They are then rotated by a tilting angle $\phi$ relative to the planar crack. The facets are smoothly integrated into the parent crack (Fig.~\ref{fig:crack_mesh_generation}c) using Boolean union operations based on Signed Distance Fields (SDFs) \cite{SDF}. Unlike traditional Boolean operations, this method creates smooth transitions between meshes. However, it can produce sharp artifacts near the edges of the meshes, which may significantly affect the calculation of configurational forces (CFs) at the crack tip. To avoid this, the smooth SDF-based union operation is applied only on the overlapping regions between the facet and the planar crack.

The mesh thickness is initially exaggerated to ensure a smoother integration of the facets into the planar crack. Then, to obtain a realistic geometric model of the crack, the mesh is shrunk by moving each vertex inward along the opposite direction of its normal vector. Further refinement is performed by applying Laplacian Smoothing with Shrinkage Reduction \cite{Vollmer1999}, uniform remeshing with PyMeshLab \cite{pymeshlab}, and Gaussian convolution using k-nearest neighbors for additional smoothing (Fig.~\ref{fig:crack_mesh_generation}d).

At each step of the mesh generation process, the model is refined using Meshfix \cite{Attene2010}, to ensure the mesh is watertight and free from defects.

\section{Dynamics-driven Mesh Relaxation: DynaMesh-R}\label{Molecular-dynamics}

To represent the coalescence of adjacent facets due to the nucleation of a type B crack,  a smooth mesh with high geometric regularity at the facet junctions is required. The previously described mesh generation method (~\ref{Model-crack-mesh-description}) can be extended to this case by directly connecting mesh nodes along the facet edges. However, this approach leads to non-smooth transitions and irregular elements, which compromise the accuracy of configurational force evaluations.

To mitigate these distortions, we developed DynaMesh-R, a mesh-relaxation method based on molecular dynamics (MD) simulations and implemented with HOOMD-blue \cite{Anderson2008,Glaser2015}. In DynaMesh-R, each mesh vertex is treated as a particle, and each edge as an elastic bond with an equilibrium length initially set to its undeformed value. This formulation preserves local geometry and promotes isotropic smoothing. The interaction between two bonded vertices $i$ and $j$ is modeled by a harmonic bond potential:

\begin{equation}\label{harm_bond_pot}
V_{ij}(r_{ij}) \;=\; \tfrac12 \, k_{ij} \bigl[r_{ij}-r_{0,ij}\bigr]^{2},
\end{equation}

where $r_{ij} = \lVert \mathbf{r}{j}-\mathbf{r}{i}\rVert$ represents the instantaneous edge length, $r_{0,ij}$ denotes the equilibrium length, and $k_{ij}$ is the force constant. The MD simulations are conducted using a time step of $0.0005$, with the Boltzmann constant $k_B$ set to $1$, the force constant $k_{ij}=300$, and equilibrium edge length $r_{0,ij}=1$.
Initially, simulations are performed under the NPT ensemble, employing a Bussi thermostat \cite{Bussi2007}, with the simulation box allowed to scale isotropically. The target pressure tensor is fixed at $1.01$, and the temperature is maintained at $1$. This NPT-based relaxation step serves as an initial smoothing of the mesh edge lengths, which initially exhibit diverse vertex spacings due to the input triangulation.
To replicate a targeted coalescence scenario, specific nodes are fixed to preserve the shape of non-critical regions, while other nodes are allowed to relax freely. Importantly, no intrinsic topological adjustments (e.g., re-meshing, edge flipping) are performed in these non-critical areas, thereby ensuring mesh quality improvements remain purely geometrical.
After 100 MD simulation iterations, the mesh undergoes a secondary relaxation phase to correct residual irregularities introduced by dynamic simulations. This phase involved numerically minimizing the energy functional defined by the surface area, $E = \int_S dA$, where $S$ is the triangulated mesh surface, and $dA$ represents an infinitesimal surface element. This relaxation models the physical evolution of a surface toward mechanical equilibrium.
Numerically, the minimization process begins by defining $\mathbf{X} = (x_1, y_1, z_1, \dots, x_n, y_n, z_n)^\top$ as the concatenated vector of vertex positions. The energy $E(\mathbf{X})$ is first minimized using gradient descent optimization with a step size $\alpha = 5$. Subsequently, to achieve a refined correction, the Hessian matrix $H_{ij} = \frac{\partial^2 E}{\partial x_{i} \partial x_{j}}$ is computed and a Newton-Raphson optimization step is applied. This two-step optimization procedure employs a gradient convergence threshold of $10^{-5}$ and a maximum iteration count of 60, and relaxation is considered complete when either criterion is met.
The initial gradient descent phase rapidly resolves significant mesh distortions induced by the MD simulation, while the subsequent Hessian-based optimization precisely refines the mesh, explicitly accounting for surface curvature and second-order geometric features such as ripples and local mesh irregularities.

\begin{figure}[htb]
    \centering
    \includegraphics[width=1.0\textwidth, keepaspectratio]{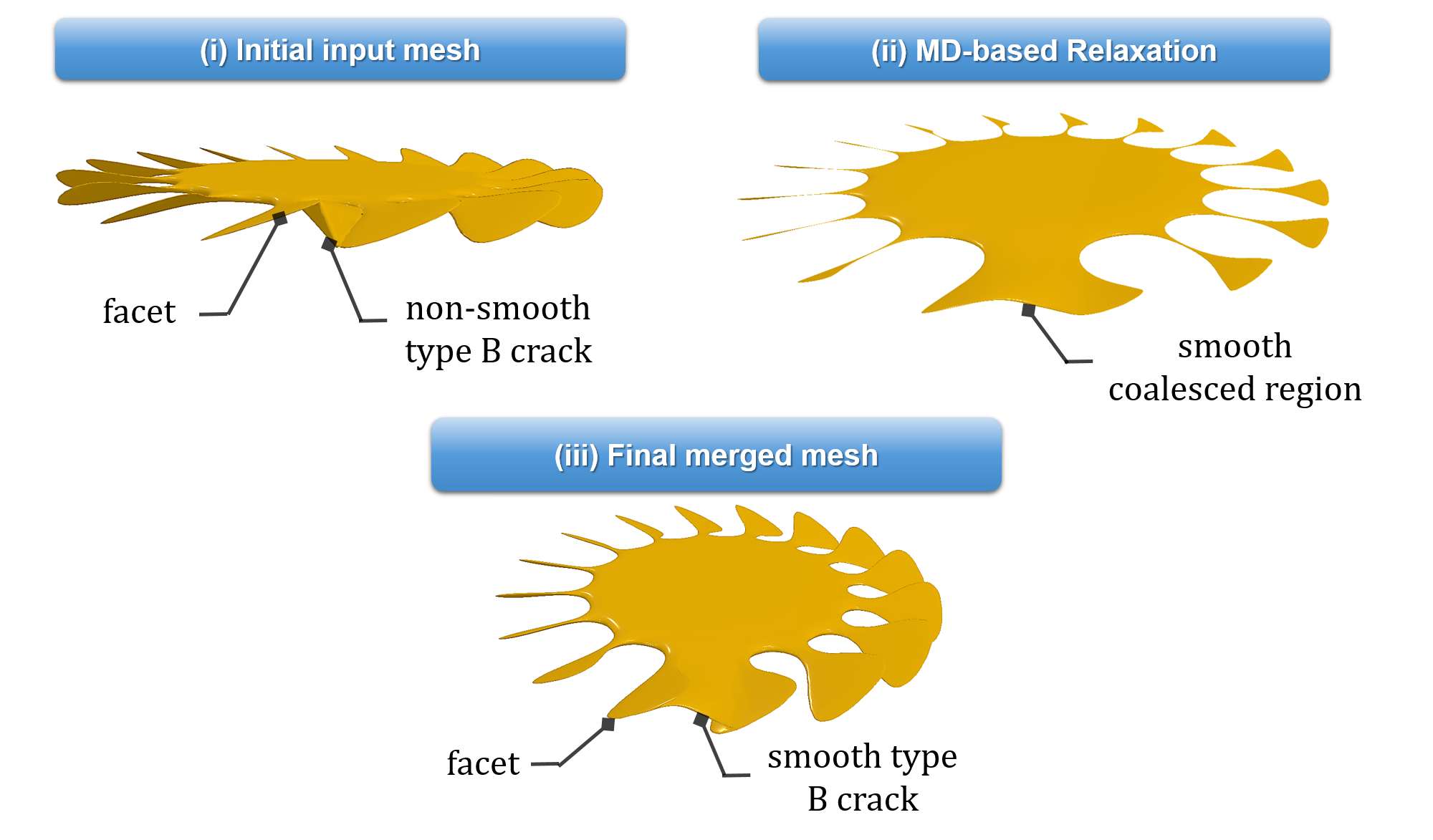}
    \caption{\textbf{Mesh processing workflow in the DynaMesh-R pipeline.}
    (i) Initial input mesh with a non-smooth junction between neighboring facets, used as the starting point for the DynaMesh-R routine. 
    (ii) Mesh after molecular dynamics–based relaxation, showing improved regularity and geometric continuity in the coalescence region. 
    (iii) Final mesh after Boolean union between a mesh without coalescence and the mesh obtained in (ii). This one is used to generate the tetrahedral mesh for configurational force analysis.
    }
    \label{fig:md_mesh_generation}
\end{figure}

Summarizing the main steps of the procedure, as illustrated in Figure~\ref{fig:md_mesh_generation}:  
(i) The mesh generated using the procedure in Section~\ref{Model-crack-mesh-description} serves as input for the MD-based relaxation routine, with neighboring facets connected through their nodes at their edges;  
(ii) An MD simulation is performed to relax and smooth the mesh, particularly in the coalesced facet region;  
(iii) The relaxed region is merged with the original mesh (non-coalesced facets) using a Boolean union operation to ensure that the overall crack shape and size are preserved. The resulting surface mesh is then used to generate a tetrahedral volume mesh for use in finite element simulations.

This procedure reduces local distortion, improves element quality, and ensures a continuous, smooth geometry at the facet coalescence region.  The resulting mesh preserves the intended crack morphology while ensuring numerical stability in the finite element simulations and accuracy in the computation of configurational forces related to type B crack formation. 

Interestingly, the procedure is general and can be applied to arbitrary meshes, regardless of how the initial geometry is generated.

\section{Effect of facet spacing and specimen size on the configurational‐force angle $\beta$} \label{Facet-spacing-beta}

We computed the normalized configurational force  \( \mathbf{F}_{\text{CNF}}/s \) during both the mode I and mode I+III loading phases, focusing on cracks with a fixed tilt angle  (\( \phi = 20^\circ \)), while (i) varying facet spacing to assess the influence of facet–facet interactions, and (ii) increasing the overall sample size. In all cases, the facet geometry and crack size remain unchanged.

\begin{figure}[htbp]
 \centering
 \includegraphics[width=1.0\textwidth, keepaspectratio]{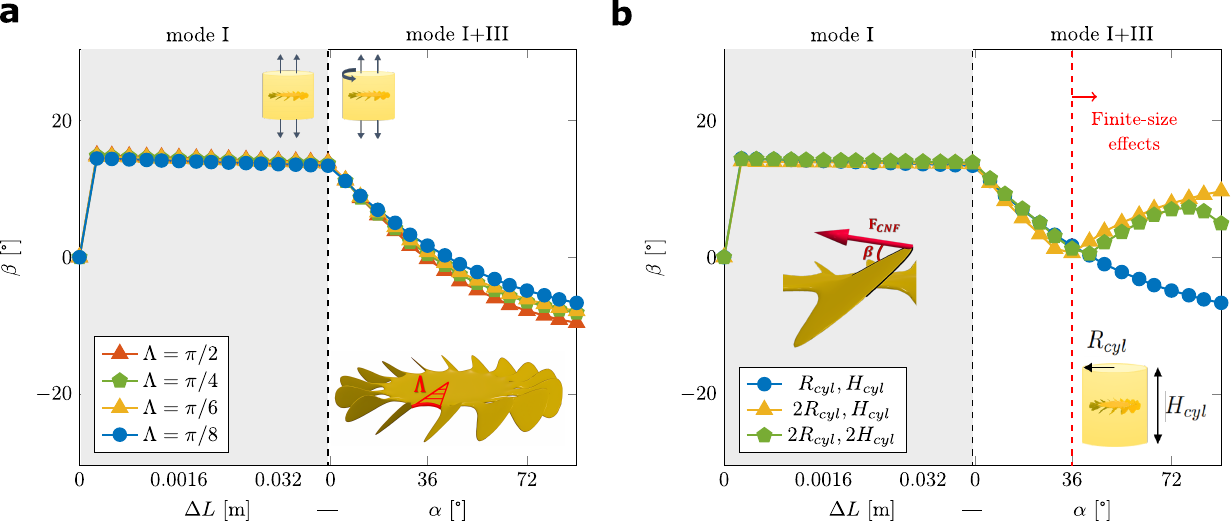}
 \caption{\textbf{Facet spacing and specimen size have limited influence on the configurational-force direction at low to intermediate torsion.} 
 (a) Evolution of the angle $\beta$—defined between the configurational force vector and the facet plane—during the mode I and mode I+III loading phases for cracks with varying facet spacing $\Lambda$.
 (b) Evolution of $\beta$ for different specimen sizes, achieved by uniformly scaling the cylinder height ($H_{\mathrm{cyl}}$) and radius ($R_{\mathrm{cyl}}$), while keeping the facet geometry and crack size fixed.
 During the mode I phase and for low to intermediate torsion angles, $\beta$ remains largely unaffected by either facet spacing or specimen size. However, for high torsion angles ($\alpha > 36^\circ$), $\beta$ increases in the larger specimens, indicating a possible finite-size effect at high twist. The red dashed vertical line marks the onset of this deviation. 
 CF values are averaged across all facets. Schematic insets illustrate the applied loading and the definition of $\beta$, $\Lambda$ and the cylinder dimensions. }
 \label{fig:beta_facets-spacing_specimen-size}
 \end{figure}

Figure~\ref{fig:beta_facets-spacing_specimen-size}a shows the evolution of the angle $\beta$—defined between the configurational force vector and the facet plane—throughout the loading path for different facet spacings $\Lambda$. During the mode I phase, $\beta$ remains nearly constant across all values of $\Lambda$, with only minimal variation. In the mode I+III phase, a slight decrease in $\beta$ is observed as $\Lambda$ decreases, though the overall differences remain small.

To assess potential finite-size effects, we repeated the analysis while increasing both the cylinder height ($H_{\mathrm{cyl}}$) and radius ($R_{\mathrm{cyl}}$), keeping the crack geometry unchanged. As shown in Fig.~\ref{fig:beta_facets-spacing_specimen-size}b, these changes in specimen size produce negligible variations in $\beta$, indicating that system-scale geometry does not significantly affect the configurational force direction.

To assess potential finite-size effects, we repeated the analysis while increasing the cylinder height ($H_{\mathrm{cyl}}$) only or together with the radius ($R_{\mathrm{cyl}}$), keeping the crack geometry unchanged. $H_{\mathrm{cyl}}$ and $R_{\mathrm{cyl}}$ are the height and radius, respectively, of the reference specimen used in the simulations. As shown in Fig.~\ref{fig:beta_facets-spacing_specimen-size}b, increasing the specimen size produces negligible changes in $\beta$ during the mode I phase and at low to intermediate torsion angles. However, for high torsion angles ($\alpha > 36^\circ$), $\beta$  increases in the larger specimens, in contrast to the continued decrease observed in the reference geometry. This trend suggests that finite-size effects may influence crack-front mechanics at large torsional loads, though they remain absent under moderate loading.

Taken together, these results demonstrate that both facet–facet interactions and finite specimen size have only a marginal influence on the angle $\beta$ over the majority of the loading path, and therefore do not account for the deviations from the LEFM-based scaling observed in Section~\ref{Scaling}.

\bibliographystyle{unsrt}  


\begin{thebibliography}{10}
\expandafter\ifx\csname url\endcsname\relax
  \def\url#1{\texttt{#1}}\fi
\expandafter\ifx\csname urlprefix\endcsname\relax\def\urlprefix{URL }\fi
\providecommand{\bibinfo}[2]{#2}
\providecommand{\eprint}[2][]{\url{#2}}

\bibitem{Buehler2006}
\bibinfo{author}{Buehler, M.~J.} \& \bibinfo{author}{Gao, H.}
\newblock \bibinfo{title}{Dynamical fracture instabilities due to local
  hyperelasticity at crack tips}.
\newblock \emph{\bibinfo{journal}{Nature}} \textbf{\bibinfo{volume}{439}},
  \bibinfo{pages}{307--310} (\bibinfo{year}{2006}).

\bibitem{RaviChandar1998}
\bibinfo{author}{Ravi-Chandar, K.}
\newblock \bibinfo{title}{Dynamic fracture of nominally brittle materials}.
\newblock \emph{\bibinfo{journal}{International Journal of Fracture}}
  \textbf{\bibinfo{volume}{90}}, \bibinfo{pages}{83--102}
  (\bibinfo{year}{1998}).

\bibitem{Fineberg1991}
\bibinfo{author}{Fineberg, J.}, \bibinfo{author}{Gross, S.~P.},
  \bibinfo{author}{Marder, M.} \& \bibinfo{author}{Swinney, H.~L.}
\newblock \bibinfo{title}{Instability in dynamic fracture}.
\newblock \emph{\bibinfo{journal}{Physical Review Letters}}
  \textbf{\bibinfo{volume}{67}}, \bibinfo{pages}{457--460}
  (\bibinfo{year}{1991}).

\bibitem{Kolvin2015}
\bibinfo{author}{Kolvin, I.}, \bibinfo{author}{Cohen, G.} \&
  \bibinfo{author}{Fineberg, J.}
\newblock \bibinfo{title}{Crack front dynamics: The interplay of singular
  geometry and crack instabilities}.
\newblock \emph{\bibinfo{journal}{Physical Review Letters}}
  \textbf{\bibinfo{volume}{114}}, \bibinfo{pages}{175501}
  (\bibinfo{year}{2015}).

\bibitem{GoldmanBoue2015b}
\bibinfo{author}{Boué, T.~G.}, \bibinfo{author}{Cohen, G.} \&
  \bibinfo{author}{Fineberg, J.}
\newblock \bibinfo{title}{Origin of the microbranching instability in rapid
  cracks}.
\newblock \emph{\bibinfo{journal}{Physical Review Letters}}
  \textbf{\bibinfo{volume}{114}}, \bibinfo{pages}{054301}
  (\bibinfo{year}{2015}).

\bibitem{Wu2025}
\bibinfo{author}{Wu, K.} \emph{et~al.}
\newblock \bibinfo{title}{Distorting crack-front geometry for enhanced
  toughness by manipulating bioinspired heterogeneity}.
\newblock \emph{\bibinfo{journal}{Nature Communications}}
  \textbf{\bibinfo{volume}{16}}, \bibinfo{pages}{1--13} (\bibinfo{year}{2025}).

\bibitem{Steinhardt2022}
\bibinfo{author}{Steinhardt, W.} \& \bibinfo{author}{Rubinstein, S.~M.}
\newblock \bibinfo{title}{How material heterogeneity creates rough fractures}.
\newblock \emph{\bibinfo{journal}{Physical Review Letters}}
  \textbf{\bibinfo{volume}{129}}, \bibinfo{pages}{128001}
  (\bibinfo{year}{2022}).

\bibitem{Steinhardt2023}
\bibinfo{author}{Steinhardt, W.} \& \bibinfo{author}{Rubinstein, S.~M.}
\newblock \bibinfo{title}{Geometric rules for the annihilation dynamics of step
  lines on fracture fronts}.
\newblock \emph{\bibinfo{journal}{Physical Review E}}
  \textbf{\bibinfo{volume}{107}}, \bibinfo{pages}{055003}
  (\bibinfo{year}{2023}).

\bibitem{Lechenault2015}
\bibinfo{author}{Lechenault, F.}, \bibinfo{author}{Sapoval, B.} \&
  \bibinfo{author}{Adda-Bedia, M.}
\newblock \bibinfo{title}{Morphology and dynamics of a crack front propagating
  in a model disordered material}.
\newblock \emph{\bibinfo{journal}{Journal of the Mechanics and Physics of
  Solids}} \textbf{\bibinfo{volume}{74}}, \bibinfo{pages}{38--48}
  (\bibinfo{year}{2015}).

\bibitem{Ramanathan1997}
\bibinfo{author}{Ramanathan, S.} \& \bibinfo{author}{Fisher, D.~S.}
\newblock \bibinfo{title}{Quasistatic crack propagation in heterogeneous
  media}.
\newblock \emph{\bibinfo{journal}{Physical Review Letters}}
  \textbf{\bibinfo{volume}{79}}, \bibinfo{pages}{873--876}
  (\bibinfo{year}{1997}).

\bibitem{Pham2016}
\bibinfo{author}{Pham, K.~H.} \& \bibinfo{author}{Ravi-Chandar, K.}
\newblock \bibinfo{title}{On the growth of cracks under mixed-mode i + iii
  loading}.
\newblock \emph{\bibinfo{journal}{International Journal of Fracture}}
  \textbf{\bibinfo{volume}{199}}, \bibinfo{pages}{105--134}
  (\bibinfo{year}{2016}).

\bibitem{Ronsin2014}
\bibinfo{author}{Ronsin, O.}, \bibinfo{author}{Caroli, C.} \&
  \bibinfo{author}{Baumberger, T.}
\newblock \bibinfo{title}{Crack front échelon instability in mixed mode
  fracture of a strongly nonlinear elastic solid}.
\newblock \emph{\bibinfo{journal}{Europhysics Letters}}
  \textbf{\bibinfo{volume}{105}}, \bibinfo{pages}{34001}
  (\bibinfo{year}{2014}).

\bibitem{Goldstein2012}
\bibinfo{author}{Goldstein, R.~V.} \& \bibinfo{author}{Osipenko, N.~M.}
\newblock \bibinfo{title}{Fracture structure near a longitudinal shear
  macrorupture}.
\newblock \emph{\bibinfo{journal}{Mechanics of Solids}}
  \textbf{\bibinfo{volume}{47}}, \bibinfo{pages}{565--574}
  (\bibinfo{year}{2012}).

\bibitem{Goldstein2014}
\bibinfo{author}{Goldstein, R.~V.} \& \bibinfo{author}{Osipenko, N.~M.}
\newblock \bibinfo{title}{Development of multiple ordered fracture in an
  elastic homogeneous, structured and layered medium}.
\newblock \emph{\bibinfo{journal}{Fatigue \& Fracture of Engineering Materials
  \& Structures}} \textbf{\bibinfo{volume}{37}}, \bibinfo{pages}{1292--1305}
  (\bibinfo{year}{2014}).

\bibitem{Pollard1982}
\bibinfo{author}{Pollard, D.~D.}, \bibinfo{author}{Segall, P.} \&
  \bibinfo{author}{Delaney, P.~T.}
\newblock \bibinfo{title}{Formation and interpretation of dilatant echelon
  cracks}.
\newblock \emph{\bibinfo{journal}{GSA Bulletin}} \textbf{\bibinfo{volume}{93}},
  \bibinfo{pages}{1291--1303} (\bibinfo{year}{1982}).

\bibitem{SOMMER1969539}
\bibinfo{author}{Sommer, E.}
\newblock \bibinfo{title}{Formation of fracture ‘lances’ in glass}.
\newblock \emph{\bibinfo{journal}{Engineering Fracture Mechanics}}
  \textbf{\bibinfo{volume}{1}}, \bibinfo{pages}{539--546}
  (\bibinfo{year}{1969}).

\bibitem{Lazarus2008}
\bibinfo{author}{Lazarus, V.}, \bibinfo{author}{Buchholz, F.-G.},
  \bibinfo{author}{Fulland, M.} \& \bibinfo{author}{Wiebesiek, J.}
\newblock \bibinfo{title}{Comparison of predictions by mode ii or mode iii
  criteria on crack front twisting in three or four point bending experiments}.
\newblock \emph{\bibinfo{journal}{International Journal of Fracture}}
  \textbf{\bibinfo{volume}{153}}, \bibinfo{pages}{141--151}
  (\bibinfo{year}{2008}).

\bibitem{Lin2010}
\bibinfo{author}{Lin, B.}, \bibinfo{author}{Mear, M.~E.} \&
  \bibinfo{author}{Ravi-Chandar, K.}
\newblock \bibinfo{title}{Criterion for initiation of cracks under mixed-mode i
  + iii loading}.
\newblock \emph{\bibinfo{journal}{International Journal of Fracture}}
  \textbf{\bibinfo{volume}{165}}, \bibinfo{pages}{175--188}
  (\bibinfo{year}{2010}).

\bibitem{Pham2014}
\bibinfo{author}{Pham, K.~H.} \& \bibinfo{author}{Ravi-Chandar, K.}
\newblock \bibinfo{title}{Further examination of the criterion for crack
  initiation under mixed-mode i+iii loading}.
\newblock \emph{\bibinfo{journal}{International Journal of Fracture}}
  \textbf{\bibinfo{volume}{189}}, \bibinfo{pages}{121--138}
  (\bibinfo{year}{2014}).

\bibitem{Wu2006}
\bibinfo{author}{Wu, R.}
\newblock \emph{\bibinfo{title}{Some Fundamental Mechanisms of Hydraulic
  Fracturing}}.
\newblock \bibinfo{type}{Doctoral dissertation}, \bibinfo{school}{Georgia
  Institute of Technology} (\bibinfo{year}{2006}).

\bibitem{ortellado2025principle}
\bibinfo{author}{Ortellado, L.}, \bibinfo{author}{Abate, A.},
  \bibinfo{author}{Santarossa, A.}, \bibinfo{author}{G{\'o}mez, L.~R.} \&
  \bibinfo{author}{P{\"o}schel, T.}
\newblock \bibinfo{title}{Principle of local symmetry in mixed-mode fracture}.
\newblock \emph{\bibinfo{journal}{Communications Physics}}
  \textbf{\bibinfo{volume}{8}}, \bibinfo{pages}{1--11} (\bibinfo{year}{2025}).

\bibitem{LEBLOND2011}
\bibinfo{author}{Leblond, J.-B.}, \bibinfo{author}{Karma, A.} \&
  \bibinfo{author}{Lazarus, V.}
\newblock \bibinfo{title}{Theoretical analysis of crack front instability in
  mode i+iii}.
\newblock \emph{\bibinfo{journal}{Journal of the Mechanics and Physics of
  Solids}} \textbf{\bibinfo{volume}{59}}, \bibinfo{pages}{1872--1887}
  (\bibinfo{year}{2011}).

\bibitem{CAMBONIE2014}
\bibinfo{author}{Cambonie, T.} \& \bibinfo{author}{Lazarus, V.}
\newblock \bibinfo{title}{Quantification of the crack fragmentation resulting
  from mode i+iii loading}.
\newblock \emph{\bibinfo{journal}{Procedia Materials Science}}
  \textbf{\bibinfo{volume}{3}}, \bibinfo{pages}{1816--1821}
  (\bibinfo{year}{2014}).
\newblock \bibinfo{note}{20th European Conference on Fracture}.

\bibitem{Vasudevan2018}
\bibinfo{author}{Vasudevan, A.~V.}
\newblock \emph{\bibinfo{title}{Deciphering triangular fracture patterns in
  PMMA: how crack fragments in mixed mode loading}}.
\newblock \bibinfo{type}{Ph.d. thesis}, \bibinfo{school}{Sorbonne Université}
  (\bibinfo{year}{2018}).

\bibitem{Lebihain2022}
\bibinfo{author}{Lebihain, M.}, \bibinfo{author}{Leblond, J.-B.} \&
  \bibinfo{author}{Ponson, L.}
\newblock \bibinfo{title}{Crack front instability in mixed-mode i+iii: The
  influence of non-singular stresses}.
\newblock \emph{\bibinfo{journal}{European Journal of Mechanics - A/Solids}}
  \textbf{\bibinfo{volume}{100}}, \bibinfo{pages}{104602}
  (\bibinfo{year}{2022}).

\bibitem{VASUDEVAN2020}
\bibinfo{author}{Vasudevan, A.}, \bibinfo{author}{Ponson, L.},
  \bibinfo{author}{Karma, A.} \& \bibinfo{author}{Leblond, J.-B.}
\newblock \bibinfo{title}{Configurational stability of a crack propagating in a
  material with mode-dependent fracture energy – part ii: Drift of fracture
  facets in mixed-mode i+ii+iii}.
\newblock \emph{\bibinfo{journal}{Journal of the Mechanics and Physics of
  Solids}} \textbf{\bibinfo{volume}{137}}, \bibinfo{pages}{103894}
  (\bibinfo{year}{2020}).

\bibitem{LEBLOND2019187}
\bibinfo{author}{Leblond, J.-B.}, \bibinfo{author}{Karma, A.},
  \bibinfo{author}{Ponson, L.} \& \bibinfo{author}{Vasudevan, A.}
\newblock \bibinfo{title}{Configurational stability of a crack propagating in a
  material with mode-dependent fracture energy - part i: Mixed-mode i+iii}.
\newblock \emph{\bibinfo{journal}{Journal of the Mechanics and Physics of
  Solids}} \textbf{\bibinfo{volume}{126}}, \bibinfo{pages}{187--203}
  (\bibinfo{year}{2019}).

\bibitem{BAHMANI2021}
\bibinfo{author}{Bahmani, A.} \emph{et~al.}
\newblock \bibinfo{title}{On the comparison of two mixed-mode i + iii
  fracture test specimens}.
\newblock \emph{\bibinfo{journal}{Engineering Fracture Mechanics}}
  \textbf{\bibinfo{volume}{241}}, \bibinfo{pages}{107434}
  (\bibinfo{year}{2021}).

\bibitem{Hodgdon1993}
\bibinfo{author}{Hodgdon, J.~A.} \& \bibinfo{author}{Sethna, J.~P.}
\newblock \bibinfo{title}{Derivation of a general three-dimensional
  crack-propagation law: A generalization of the principle of local symmetry}.
\newblock \emph{\bibinfo{journal}{Physical Review B}}
  \textbf{\bibinfo{volume}{47}}, \bibinfo{pages}{4831--4840}
  (\bibinfo{year}{1993}).

\bibitem{Molnar2024}
\bibinfo{author}{Molnár, G.}, \bibinfo{author}{Doitrand, A.} \&
  \bibinfo{author}{Lazarus, V.}
\newblock \bibinfo{title}{Phase-field simulation and coupled criterion link
  echelon cracks to internal length in antiplane shear}.
\newblock \emph{\bibinfo{journal}{Journal of the Mechanics and Physics of
  Solids}} \textbf{\bibinfo{volume}{188}}, \bibinfo{pages}{105675}
  (\bibinfo{year}{2024}).

\bibitem{Bouchbinder2009}
\bibinfo{author}{Bouchbinder, E.}, \bibinfo{author}{Livne, A.} \&
  \bibinfo{author}{Fineberg, J.}
\newblock \bibinfo{title}{The 1/r singularity in weakly nonlinear fracture
  mechanics}.
\newblock \emph{\bibinfo{journal}{Journal of the Mechanics and Physics of
  Solids}} \textbf{\bibinfo{volume}{57}}, \bibinfo{pages}{1568}
  (\bibinfo{year}{2009}).

\bibitem{Bouchbinder2009b}
\bibinfo{author}{Bouchbinder, E.}, \bibinfo{author}{Livne, A.} \&
  \bibinfo{author}{Fineberg, J.}
\newblock \bibinfo{title}{Weakly nonlinear fracture mechanics: experiments and
  theory}.
\newblock \emph{\bibinfo{journal}{International Journal of Fracture}}
  \textbf{\bibinfo{volume}{161}}, \bibinfo{pages}{1--20}
  (\bibinfo{year}{2009}).

\bibitem{Bouchbinder2008}
\bibinfo{author}{Bouchbinder, E.}, \bibinfo{author}{Livne, A.} \&
  \bibinfo{author}{Fineberg, J.}
\newblock \bibinfo{title}{Weakly nonlinear theory of dynamic fracture}.
\newblock \emph{\bibinfo{journal}{Physical Review Letters}}
  \textbf{\bibinfo{volume}{101}}, \bibinfo{pages}{264302}
  (\bibinfo{year}{2008}).

\bibitem{GoldmanBoue2015}
\bibinfo{author}{Bou{\'e}, T.~G.}, \bibinfo{author}{Harpaz, R.},
  \bibinfo{author}{Fineberg, J.} \& \bibinfo{author}{Bouchbinder, E.}
\newblock \bibinfo{title}{Failing softly: a fracture theory of
  highly-deformable materials}.
\newblock \emph{\bibinfo{journal}{Soft Matter}} \textbf{\bibinfo{volume}{11}},
  \bibinfo{pages}{3812--3821} (\bibinfo{year}{2015}).

\bibitem{Livne2010}
\bibinfo{author}{Livne, A.}, \bibinfo{author}{Bouchbinder, E.} \&
  \bibinfo{author}{Fineberg, J.}
\newblock \bibinfo{title}{The near-tip fields of fast cracks}.
\newblock \emph{\bibinfo{journal}{Science}} \textbf{\bibinfo{volume}{327}},
  \bibinfo{pages}{1359--1363} (\bibinfo{year}{2010}).

\bibitem{Livne2008}
\bibinfo{author}{Livne, A.}, \bibinfo{author}{Bouchbinder, E.} \&
  \bibinfo{author}{Fineberg, J.}
\newblock \bibinfo{title}{Breakdown of linear elastic fracture mechanics near
  the tip of a rapid crack}.
\newblock \emph{\bibinfo{journal}{Physical Review Letters}}
  \textbf{\bibinfo{volume}{101}}, \bibinfo{pages}{264301}
  (\bibinfo{year}{2008}).

\bibitem{Wei2024}
\bibinfo{author}{Wei, X.}, \bibinfo{author}{Li, C.}, \bibinfo{author}{McCarthy,
  C.} \& \bibinfo{author}{Kolinski, J.~M.}
\newblock \bibinfo{title}{Complexity of crack front geometry enhances toughness
  of brittle solids}.
\newblock \emph{\bibinfo{journal}{Nature Physics}}
  \textbf{\bibinfo{volume}{20}}, \bibinfo{pages}{1009--1014}
  (\bibinfo{year}{2024}).

\bibitem{Pons2010}
\bibinfo{author}{Pons, A.~J.} \& \bibinfo{author}{Karma, A.}
\newblock \bibinfo{title}{Helical crack-front instability in mixed-mode
  fracture}.
\newblock \emph{\bibinfo{journal}{Nature 2010 464:7285}}
  \textbf{\bibinfo{volume}{464}}, \bibinfo{pages}{85--89}
  (\bibinfo{year}{2010}).

\bibitem{Chen2015}
\bibinfo{author}{Chen, C.-H.} \emph{et~al.}
\newblock \bibinfo{title}{Crack front segmentation and facet coarsening in
  mixed-mode fracture}.
\newblock \emph{\bibinfo{journal}{Phys. Rev. Lett.}}
  \textbf{\bibinfo{volume}{115}}, \bibinfo{pages}{265503}
  (\bibinfo{year}{2015}).

\bibitem{Henry_2016}
\bibinfo{author}{Henry, H.}
\newblock \bibinfo{title}{Crack front instabilities under mixed mode loading in
  three dimensions}.
\newblock \emph{\bibinfo{journal}{Europhysics Letters}}
  \textbf{\bibinfo{volume}{114}}, \bibinfo{pages}{66001}
  (\bibinfo{year}{2016}).

\bibitem{Leblond2015}
\bibinfo{author}{Leblond, J.-B.}, \bibinfo{author}{Lazarus, V.} \&
  \bibinfo{author}{Karma, A.}
\newblock \bibinfo{title}{Multiscale cohesive zone model for propagation of
  segmented crack fronts in mode i+ iii fracture}.
\newblock \emph{\bibinfo{journal}{International Journal of Fracture}}
  \textbf{\bibinfo{volume}{191}}, \bibinfo{pages}{167--189}
  (\bibinfo{year}{2015}).

\bibitem{Lazarus2020}
\bibinfo{author}{Lazarus, V.}, \bibinfo{author}{Prabel, B.},
  \bibinfo{author}{Cambonie, T.} \& \bibinfo{author}{Leblond, J.}
\newblock \bibinfo{title}{Mode i+ iii multiscale cohesive zone model with facet
  coarsening and overlap: Solutions and applications to facet orientation and
  toughening}.
\newblock \emph{\bibinfo{journal}{Journal of the Mechanics and Physics of
  Solids}} \textbf{\bibinfo{volume}{141}}, \bibinfo{pages}{104007}
  (\bibinfo{year}{2020}).

\bibitem{Hattali2021}
\bibinfo{author}{Hattali, M.}, \bibinfo{author}{Cambonie, T.} \&
  \bibinfo{author}{Lazarus, V.}
\newblock \bibinfo{title}{Toughening induced by the formation of facets in mode
  i+ iii brittle fracture: experiments versus a two-scale cohesive zone model}.
\newblock \emph{\bibinfo{journal}{Journal of the Mechanics and Physics of
  Solids}} \textbf{\bibinfo{volume}{156}}, \bibinfo{pages}{104596}
  (\bibinfo{year}{2021}).

\bibitem{Pandolfi2012}
\bibinfo{author}{Pandolfi, A.} \& \bibinfo{author}{Ortiz, M.}
\newblock \bibinfo{title}{An eigenerosion approach to brittle fracture}.
\newblock \emph{\bibinfo{journal}{International Journal for Numerical Methods
  in Engineering}} \textbf{\bibinfo{volume}{92}}, \bibinfo{pages}{694--714}
  (\bibinfo{year}{2012}).

\bibitem{Cherepanov1967}
\bibinfo{author}{Cherepanov, G.~P.}
\newblock \bibinfo{title}{Crack propagation in continuous media}.
\newblock \emph{\bibinfo{journal}{Journal of Applied Mathematics and
  Mechanics}} \textbf{\bibinfo{volume}{31}}, \bibinfo{pages}{503--512}
  (\bibinfo{year}{1967}).

\bibitem{Rice1968}
\bibinfo{author}{Rice, J.~R.}
\newblock \bibinfo{title}{A path independent integral and the approximate
  analysis of strain concentration by notches and cracks}.
\newblock \emph{\bibinfo{journal}{Journal of Applied Mechanics}}
  \textbf{\bibinfo{volume}{35}}, \bibinfo{pages}{379} (\bibinfo{year}{1968}).

\bibitem{Schmitz2023}
\bibinfo{author}{Schmitz, K.} \& \bibinfo{author}{Ricoeur, A.}
\newblock \bibinfo{title}{Theoretical and computational aspects of
  configurational forces in three-dimensional crack problems}.
\newblock \emph{\bibinfo{journal}{International Journal of Solids and
  Structures}} \textbf{\bibinfo{volume}{282}}, \bibinfo{pages}{112456}
  (\bibinfo{year}{2023}).

\bibitem{Moreno-Mateos2024b}
\bibinfo{author}{Moreno-Mateos, M.~A.} \& \bibinfo{author}{Steinmann, P.}
\newblock \bibinfo{title}{Configurational force method enables fracture
  assessment in soft materials}.
\newblock \emph{\bibinfo{journal}{Journal of the Mechanics and Physics of
  Solids}} \textbf{\bibinfo{volume}{186}}, \bibinfo{pages}{105602}
  (\bibinfo{year}{2024}).

\bibitem{Serrao2025b}
\bibinfo{author}{Serrao, P.~H.} \& \bibinfo{author}{Kozinov, S.}
\newblock \bibinfo{title}{Evaluation of configurational/material forces in
  strain gradient elasticity theory}.
\newblock \emph{\bibinfo{journal}{Mechanics of Materials}}
  \textbf{\bibinfo{volume}{203}}, \bibinfo{pages}{105240}
  (\bibinfo{year}{2025}).

\bibitem{Goda2016}
\bibinfo{author}{Goda, I.}, \bibinfo{author}{Ganghoffer, J.~F.} \&
  \bibinfo{author}{Maurice, G.}
\newblock \bibinfo{title}{Combined bone internal and external remodeling based
  on eshelby stress}.
\newblock \emph{\bibinfo{journal}{International Journal of Solids and
  Structures}} \textbf{\bibinfo{volume}{94-95}}, \bibinfo{pages}{138--157}
  (\bibinfo{year}{2016}).

\bibitem{Braun1997}
\bibinfo{author}{Braun, M.}
\newblock \bibinfo{title}{Configurational forces induced by finite-element
  discretization}.
\newblock \emph{\bibinfo{journal}{Proceedings of the Estonian Academy of
  Sciences : Physics, Mathematics}} \textbf{\bibinfo{volume}{46}},
  \bibinfo{pages}{24--31} (\bibinfo{year}{1997}).

\bibitem{LAZARUS2001}
\bibinfo{author}{Lazarus, V.}, \bibinfo{author}{Leblond, J.-B.} \&
  \bibinfo{author}{Mouchrif, S.-E.}
\newblock \bibinfo{title}{Crack front rotation and segmentation in mixed mode
  i+iii or i+ii+iii. part i: Calculation of stress intensity factors}.
\newblock \emph{\bibinfo{journal}{Journal of the Mechanics and Physics of
  Solids}} \textbf{\bibinfo{volume}{49}}, \bibinfo{pages}{1399--1420}
  (\bibinfo{year}{2001}).

\bibitem{Santarossa2023}
\bibinfo{author}{Santarossa, A.}, \bibinfo{author}{Ortellado, L.},
  \bibinfo{author}{Sack, A.}, \bibinfo{author}{Gómez, L.~R.} \&
  \bibinfo{author}{Pöschel, T.}
\newblock \bibinfo{title}{A device for studying fluid-induced cracks under
  mixed-mode loading conditions using x-ray tomography}.
\newblock \emph{\bibinfo{journal}{Review of Scientific Instruments}}
  \textbf{\bibinfo{volume}{94}} (\bibinfo{year}{2023}).

\bibitem{VANOTTERLOO201686}
\bibinfo{author}{{van Otterloo}, J.} \& \bibinfo{author}{Cruden, A.~R.}
\newblock \bibinfo{title}{Rheology of pig skin gelatine: Defining the elastic
  domain and its thermal and mechanical properties for geological analogue
  experiment applications}.
\newblock \emph{\bibinfo{journal}{Tectonophysics}}
  \textbf{\bibinfo{volume}{683}}, \bibinfo{pages}{86--97}
  (\bibinfo{year}{2016}).

\bibitem{Eshelby1951}
\bibinfo{author}{Eshelby, J.~D.}
\newblock \bibinfo{title}{The force on an elastic singularity}.
\newblock \emph{\bibinfo{journal}{Philosophical Transactions of the Royal
  Society of London}} \textbf{\bibinfo{volume}{244}} (\bibinfo{year}{1951}).

\bibitem{Kienzler1997}
\bibinfo{author}{Kienzler, R.} \& \bibinfo{author}{Herrmann, G.}
\newblock \bibinfo{title}{On the properties of the eshelby tensor}.
\newblock \emph{\bibinfo{journal}{Acta Mechanica}}
  \textbf{\bibinfo{volume}{125}}, \bibinfo{pages}{73--91}
  (\bibinfo{year}{1997}).

\bibitem{Eshelby1975}
\bibinfo{author}{Eshelby, J.~D.}
\newblock \bibinfo{title}{The elastic energy-momentum tensor}.
\newblock \emph{\bibinfo{journal}{Journal of Elasticity}}
  \textbf{\bibinfo{volume}{5}}, \bibinfo{pages}{321--335}
  (\bibinfo{year}{1975}).

\bibitem{Eshelby1999}
\bibinfo{author}{Eshelby, J.~D.}
\newblock \bibinfo{title}{Energy relations and the energy-momentum tensor in
  continuum mechanics}.
\newblock \emph{\bibinfo{journal}{Fundamental Contributions to the Continuum
  Theory of Evolving Phase Interfaces in Solids}} \bibinfo{pages}{82--119}
  (\bibinfo{year}{1999}).

\bibitem{Steinmann2001}
\bibinfo{author}{Steinmann, P.}, \bibinfo{author}{Ackermann, D.} \&
  \bibinfo{author}{Barth, F.}
\newblock \bibinfo{title}{Application of material forces to hyperelastostatic
  fracture mechanics. ii. computational setting}.
\newblock \emph{\bibinfo{journal}{International Journal of Solids and
  Structures}} \textbf{\bibinfo{volume}{38}}, \bibinfo{pages}{5509--5526}
  (\bibinfo{year}{2001}).

\bibitem{Denzer2003}
\bibinfo{author}{Denzer, R.}, \bibinfo{author}{Barth, F.~J.} \&
  \bibinfo{author}{Steinmann, P.}
\newblock \bibinfo{title}{Studies in elastic fracture mechanics based on the
  material force method}.
\newblock \emph{\bibinfo{journal}{International Journal for Numerical Methods
  in Engineering}} \textbf{\bibinfo{volume}{58}}, \bibinfo{pages}{1817--1835}
  (\bibinfo{year}{2003}).

\bibitem{Logg2012}
\bibinfo{author}{Logg, A.}, \bibinfo{author}{Mardal, K.-A.} \&
  \bibinfo{author}{Wells, G.}
\newblock \emph{\bibinfo{title}{Automated Solution of Differential Equations by
  the Finite Element Method}}, vol.~\bibinfo{volume}{84}
  (\bibinfo{publisher}{Springer Berlin Heidelberg}, \bibinfo{year}{2012}).

\bibitem{murakami1985}
\bibinfo{author}{Murakami, Y.}
\newblock \bibinfo{title}{Analysis of stress intensity factors of modes i, ii
  and iii for inclined surface cracks of arbitrary shape}.
\newblock \emph{\bibinfo{journal}{Engineering Fracture Mechanics}}
  \textbf{\bibinfo{volume}{22}}, \bibinfo{pages}{101--114}
  (\bibinfo{year}{1985}).

\bibitem{cotterell1980}
\bibinfo{author}{Cotterell, B.} \& \bibinfo{author}{Rice, J.~R.}
\newblock \bibinfo{title}{Slightly curved or kinked cracks}.
\newblock \emph{\bibinfo{journal}{International journal of fracture}}
  \textbf{\bibinfo{volume}{16}}, \bibinfo{pages}{155--169}
  (\bibinfo{year}{1980}).

\bibitem{thomas2017quantification}
\bibinfo{author}{Thomas, R.~N.}, \bibinfo{author}{Paluszny, A.} \&
  \bibinfo{author}{Zimmerman, R.~W.}
\newblock \bibinfo{title}{Quantification of fracture interaction using stress
  intensity factor variation maps}.
\newblock \emph{\bibinfo{journal}{Journal of Geophysical Research: Solid
  Earth}} \textbf{\bibinfo{volume}{122}}, \bibinfo{pages}{7698--7717}
  (\bibinfo{year}{2017}).

\bibitem{Li1985}
\bibinfo{author}{Li, F.}, \bibinfo{author}{Shih, C.} \&
  \bibinfo{author}{Needleman, A.}
\newblock \bibinfo{title}{A comparison of methods for calculating energy
  release rates}.
\newblock \emph{\bibinfo{journal}{Engineering Fracture Mechanics}}
  \textbf{\bibinfo{volume}{21}}, \bibinfo{pages}{405--421}
  (\bibinfo{year}{1985}).

\bibitem{Moreno-Mateos2025}
\bibinfo{author}{Moreno-Mateos, M.~A.}, \bibinfo{author}{Wiesheier, S.},
  \bibinfo{author}{Esmaeili, A.}, \bibinfo{author}{Hossain, M.} \&
  \bibinfo{author}{Steinmann, P.}
\newblock \bibinfo{title}{Biaxial characterization of soft elastomers:
  experiments and data-adaptive configurational forces for fracture}
  (\bibinfo{year}{2025}).
\newblock \eprint{2505.20244}.

\bibitem{Bishara2024}
\bibinfo{author}{Bishara, D.} \& \bibinfo{author}{Li, S.}
\newblock \bibinfo{title}{A material energy–momentum flux-driven phase field
  fracture mechanics model}.
\newblock \emph{\bibinfo{journal}{Computer Methods in Applied Mechanics and
  Engineering}} \textbf{\bibinfo{volume}{425}}, \bibinfo{pages}{116920}
  (\bibinfo{year}{2024}).

\bibitem{Moreno-Mateos2024a}
\bibinfo{author}{Moreno-Mateos, M.~A.}, \bibinfo{author}{Mehnert, M.} \&
  \bibinfo{author}{Steinmann, P.}
\newblock \bibinfo{title}{Electro-mechanical actuation modulates fracture
  performance of soft dielectric elastomers}.
\newblock \emph{\bibinfo{journal}{International Journal of Engineering
  Science}} \textbf{\bibinfo{volume}{195}}, \bibinfo{pages}{104008}
  (\bibinfo{year}{2024}).

\bibitem{Steinmann2023}
\bibinfo{author}{Steinmann, P.}, \bibinfo{author}{de~Villiers, A.},
  \bibinfo{author}{McBride, A.} \& \bibinfo{author}{Javili, A.}
\newblock \bibinfo{title}{Configurational peridynamics}.
\newblock \emph{\bibinfo{journal}{Mechanics of Materials}}
  \textbf{\bibinfo{volume}{185}}, \bibinfo{pages}{104751}
  (\bibinfo{year}{2023}).

\bibitem{SDF}
\bibinfo{author}{Jones, M.}, \bibinfo{author}{Baerentzen, J.} \&
  \bibinfo{author}{Sramek, M.}
\newblock \bibinfo{title}{3d distance fields: a survey of techniques and
  applications}.
\newblock \emph{\bibinfo{journal}{IEEE Transactions on Visualization and
  Computer Graphics}} \textbf{\bibinfo{volume}{12}}, \bibinfo{pages}{581--599}
  (\bibinfo{year}{2006}).

\bibitem{Vollmer1999}
\bibinfo{author}{Vollmer, J.}, \bibinfo{author}{Mencl, R.} \&
  \bibinfo{author}{M{\"u}ller, H.}
\newblock \bibinfo{title}{Improved laplacian smoothing of noisy surface
  meshes}.
\newblock \emph{\bibinfo{journal}{Computer Graphics Forum}}
  \textbf{\bibinfo{volume}{18}}, \bibinfo{pages}{131--138}
  (\bibinfo{year}{1999}).

\bibitem{pymeshlab}
\bibinfo{author}{Muntoni, A.} \& \bibinfo{author}{Cignoni, P.}
\newblock \bibinfo{title}{{PyMeshLab}} (\bibinfo{year}{2021}).

\bibitem{Attene2010}
\bibinfo{author}{Attene, M.}
\newblock \bibinfo{title}{A lightweight approach to repairing digitized polygon
  meshes}.
\newblock \emph{\bibinfo{journal}{The Visual Computer}}
  \textbf{\bibinfo{volume}{26}}, \bibinfo{pages}{1393--1406}
  (\bibinfo{year}{2010}).

\bibitem{Anderson2008}
\bibinfo{author}{Anderson, J.~A.}, \bibinfo{author}{Lorenz, C.~D.} \&
  \bibinfo{author}{Travesset, A.}
\newblock \bibinfo{title}{General purpose molecular dynamics simulations fully
  implemented on graphics processing units}.
\newblock \emph{\bibinfo{journal}{Journal of Computational Physics}}
  \textbf{\bibinfo{volume}{227}}, \bibinfo{pages}{5342--5359}
  (\bibinfo{year}{2008}).

\bibitem{Glaser2015}
\bibinfo{author}{Glaser, J.} \emph{et~al.}
\newblock \bibinfo{title}{Strong scaling of general-purpose molecular dynamics
  simulations on gpus}.
\newblock \emph{\bibinfo{journal}{Computer Physics Communications}}
  \textbf{\bibinfo{volume}{192}}, \bibinfo{pages}{97--107}
  (\bibinfo{year}{2015}).

\bibitem{Bussi2007}
\bibinfo{author}{Bussi, G.}, \bibinfo{author}{Donadio, D.} \&
  \bibinfo{author}{Parrinello, M.}
\newblock \bibinfo{title}{Canonical sampling through velocity rescaling}.
\newblock \emph{\bibinfo{journal}{The Journal of Chemical Physics}}
  \textbf{\bibinfo{volume}{126}}, \bibinfo{pages}{014101}
  (\bibinfo{year}{2007}).

\end{thebibliography}

\end{document}